\documentclass[conference]{IEEEtran}

\usepackage{pgf, tikz}
\usepackage{calc}
\usetikzlibrary{calc}
\usetikzlibrary{arrows}
\definecolor{mycolor1}{rgb}{0.10000,0.60000,0.30000}%
\usepackage{comment}

\usepackage{cancel}
\usepackage[normalem]{ulem}
\usepackage{xcolor}

\newcommand{\stkout}[1]{\ifmmode\text{\sout{\ensuremath{#1}}}\else\sout{#1}\fi}

\definecolor{mycolor1}{rgb}{0.10000,0.60000,0.30000}%
\definecolor{mypink2}{RGB}{219, 48, 122}
\usepackage{amssymb}
\usepackage{amsmath}
\usepackage{color}
\usepackage{bbm}
\usepackage[mode=buildnew]{standalone}
\usepackage{research5IEEE}

\newfont{\bbb}{msbm10 scaled 500}

\newfont{\bb}{msbm10 scaled 1100}

\newcommand{\xv}{{\bf x}}

\newcommand{\CDc}{{\cal{CD}}}
\newcommand{\Cc}{{\cal C}}
\newcommand{\Dc}{{\cal D}}

\newcommand{\Sc}{{\cal S}}
\newcommand{\Tc}{{\cal T}}

\newcommand{\Wc}{{\cal W}}

\newcommand{\Xc}{{\cal X}}
\newcommand{\Yc}{{\cal Y}}
\newcommand{\Zc}{{\cal Z}}

\renewcommand{\xv}{\pmb{x}}

\usepackage{color}
\usepackage{cite}
\usepackage{graphics}
\usepackage{tikz}
\usepackage{pgfplots}
\usepackage{stfloats}

\usepackage{float}
\usepackage{placeins}
\usetikzlibrary{shapes,  arrows,  decorations.markings,  arrows.meta}
\usetikzlibrary{patterns} 
\allowdisplaybreaks
\usepackage{epsfig}

\usepackage{rotating}
\usepackage{amsmath}
\usepackage{amsfonts}
\usepackage{url}

\usepackage[caption=false, font=footnotesize]{subfig}
\usepackage{epstopdf}
\usepackage{amsthm}
\usepackage{mathtools}

\usepackage{algorithm}
\usepackage[noend]{algpseudocode}

\usepackage{array}

\usepackage{cite}

\newtheorem{theorem}{Theorem}
\newtheorem{example}{Example}
\newtheorem{definition}{Definition}

\newtheorem{corollary}{Corollary}
\newtheorem{remark}{Remark}
\newcommand{\mw}[1]{{\color{black}#1}}
\definecolor{blue-green}{rgb}{0, 0.6, 0.8}

\newcommand{\D}{\mathsf{D}}

\definecolor{dgreen}{rgb}{0, 0.8, 0.4}
\newcommand{\lo}[1]{{\color{black}#1}}
\newcommand{\sh}[1]{}

\newcommand{\R}{\mathsf{R}}

\newcommand*{\colorboxed}{}
\def\colorboxed#1#{%
	\colorboxedAux{#1}%
}
\newcommand*{\colorboxedAux}[3]{%
	\begingroup
	\colorlet{cb@saved}{.}%
	\color#1{#2}%
	\boxed{%
		\color{cb@saved}%
		#3%
	}%
	\endgroup
}

\allowdisplaybreaks[4]
\pgfplotsset{compat=1.16} 
\title{Coding for Sensing: An Improved  Scheme for Integrated  Sensing and Communication  over MACs}
\begin{document}
	\author{
		\IEEEauthorblockN{Mehrasa Ahmadipour\IEEEauthorrefmark{1}, Mich\`ele Wigger\IEEEauthorrefmark{1},  and Mari Kobayashi\IEEEauthorrefmark{2} 
		}
		\IEEEauthorblockA{\small\IEEEauthorrefmark{1} LTCI Telecom Paris, IP Paris, 91120 Palaiseau, France, Emails:
			\url{{mehrasa.ahmadipour,michele.wigger}@telecom-paris.fr}}
		\IEEEauthorblockA{\small\IEEEauthorrefmark{2} Technical University of Munich, Munich, Germany,  Email: mari.kobayashi@tum.de}	}
	
	\maketitle
	\begin{abstract} 
		A memoryless state-dependent multiple-access channel (MAC) is considered, where two transmitters wish to convey their messages to a single receiver while simultaneously sensing (estimating) the respective states via generalized feedbacks. For this channel, an  improved inner bound is provided on  the \emph{fundamental rate-distortions tradeoff} which characterizes the communication rates the transmitters can  achieve while  simultaneously ensuring that their  state-estimates satisfy  desired distortion criteria. The  new inner bound is based on a scheme where each transmitter codes over the generalized feedback so as to improve the state estimation  at the other transmitter. This is in contrast to the  schemes proposed for point-to-point and broadcast channels where coding is used only for the transmission of messages and the optimal estimators operate on a symbol-by-symbol basis on the sequences of channel inputs and feedback outputs.
	\end{abstract}
	
	\IEEEpeerreviewmaketitle
	\section{Introduction}
	
	In various demanding applications such as smart cities and autonomous driving, terminals have to communicate data to other terminals while at the same time also to sense the environment for changes in locations, shapes, and status characteristics of static or locomoting \mw{objects.} This integrated sensing and communication scenarios have recently received lots of attention from the communications and signal processing communities \cite{sturm2011waveform,
		bliss2014cooperative,
		chiriyath2016inner,
		paul2017survey,
		kumari2017performance,
		kumari2018ieee,
		Dokhanchi2019RadarCom,
		zheng2019radar,
		liu2020joint,
		gaudio2020effectiveness,
		Kumari2021MW-JCR,
		liu2022CramerRao} and first information-theoretic studies were presented in \cite{kobayashi2018joint,kobayashi2019joint, Mehrasa2020BC}. Specifically, \cite{kobayashi2018joint, Mehrasa2020BC} identify the optimal tradeoff between the set of achievable data rates and distortions of the state estimates that can be attained over  state-dependent point-to-point (P2P) or degraded broadcast channels (BCs) with generalized feedback. Inner and outer bounds on this tradeoff for general BCs were  proposed in \cite{Mehrasa2020BC}. In \cite{kobayashi2018joint,Mehrasa2020BC} it was further established that the transmitter's optimal estimators   in the P2P and BC setup are symbol-wise estimators applied to the sequences of the transmitter's channel inputs and feedback outputs. As a consequence, the sensing performance of these systems depends only on the distribution of the input symbols but not on the applied coding schemes. 
	
	The situation is different on the multiaccess channel, where basing the estimator only on the sequence of inputs and feedback outputs at a transmitter is suboptimal. In \cite{kobayashi2019joint} it was noticed that an estimator that bases its decision also on the codewords decoded at a transmitter can improve estimation performance. In this paper, we show that further improvement is possible if  each transmitter uses coding to convey information related to its own observed generalized feedback signal to the other transmitter. In some sense, this is the first information-theoretic completely integrated sensing and communication scheme because coding is not only used to improve data communication but also to improve sensing performance at the terminals. Both our scheme and the scheme in \cite{kobayashi2019joint} are built on  Willem's scheme for the MAC with generalized feedback \cite{willems1983achievable}. 	
	
	A related idea was previously used in \cite{Lapidoth2013MAC,Lapidoth2013MAC2} for the state-dependent  \mw{MAC, where the transmitters compress and transmit their state information to the receiver. In their setup, the transmission of the state is beneficial over pure data transmission because it helps the receiver to decode the data.} 
	In our work here, each transmitter compresses and transmits information about its feedback signal to provide state-information to the other transmitter that is not available from its own feedback. 
	
	The simultaneous state and data communication
	problem as studied in   \cite{kim2008state,zhang2011joint,choudhuri2013causal, Shraga,Sibi,Joudeh2021JBinaryDetect} is also related to our integrated communication and sensing problem. The difference between these works and the present paper is that in joint communication of data and states the state sequences(s) are available at the transmitter(s) and have to be estimated at the receiver(s). 
	%
	


	\textit{Notations:}
	We use calligraphic letters to denote sets,  e.g.,  $\Xc$. 
	Random variables are denoted by uppercase letters,  e.g.,  $X$,  and their realizations by lowercase letters,  e.g.,  $x$.
	For vectors, we use boldface notation,  i.e.,  lower case boldface letters such as $\xv$ for deterministic vectors.
	
	\mw{For positive integers $n$, we  use $[1:n]$ to denote the set $\{1, \cdots, n\}$,} and  $X^n$ for the tuple of random variables $(X_1, \cdots, X_n)$. 
	We abbreviate \emph{independent and identically distributed} as \emph{i.i.d.} {and \emph{probability mass function} as \emph{pmf}.} Logarithms are taken with respect to base $2$. For an index $k\in\{1,2\}$, we define $\bar{k}:=3-k$ \mw{and for an event $\mathcal{A}$ we denote its complement by $\bar{\mathcal{A}}$.} 
	Moreover, $\mathbbm{1}\{\cdot\}$ denotes the indicator function.
	\section{System Model}

\begin{figure}[h!]
		\includegraphics[scale=0.72]{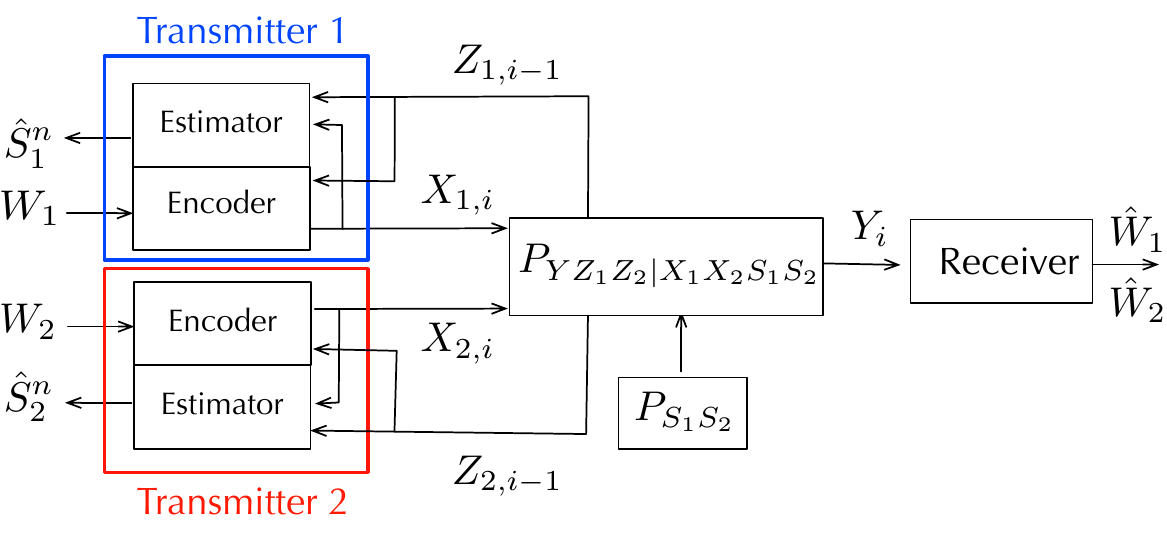}
		\vspace{-4mm}
		\caption{State-dependent discrete memoryless multiaccess channel with sensing at the transmitters.}
		\label{fig:ModelMAC}
		\vspace{-1mm}
	\end{figure}

	Consider the two-transmitter (Tx) single-receiver (Rx) multiaccess channel scenario  in Fig.~\ref{fig:ModelMAC}.  The model consists of a two-dimensional memoryless state sequence $\{(S_{1, i},  S_{2, i})\}_{i\geq 1}$ whose samples at any given time $i$ are distributed according to a given joint law $P_{S_1S_2}$ over the state alphabets $\Sc_1\times \Sc_2$. 
	Given that at time-$i$ Tx 1 sends input $X_{1,i}=x_1$ and Tx~2 input $X_{2,i}=x_2$ and given  state realizations $S_{1,i}=s_{1}$ and $S_{2,i}=s_{2}$,  the Rx's  time-$i$ output $Y_{i}$  and the Txs'  feedback signals $Z_{1,i}$ and $Z_{2,i}$ are distributed according to the stationary channel transition law $ P_{YZ_1Z_2|S_1S_2X_1X_2}(\cdot,\cdot,\cdot|s_1,s_2,x_1,x_2)$. Input and output 
	alphabets $\Xc_1, \Xc_2,  \Yc,  \Zc_1,  \Zc_2, \Sc_1, \Sc_2$ are assumed  finite. 
	
	A $(2^{n\R_1}, 2^{n\R_2},  n)$ code  consists  of
	\begin{enumerate}
		\item two message sets $\Wc_1= [1:2^{n\R_1}]$, and $\Wc_2= [1:2^{n\R_2}]$;
		\item a sequence of encoding functions $\Omega_{k,i}\colon \Wc_k \times \Zc_k^{{i-1} }\to \Xc_k$,  for $i=1, 2, \ldots, n$ and $k=1,2$; 
		\item  a decoding function $g \colon \Yc^n \to \Wc_1\times \Wc_2$; 
		\item for each $k=1, 2$  a state estimator  $\phi_k \colon \Xc_k^n \times \Zc_k^n \to \hat{\Sc}_k^n$,  where  $\hat{\Sc}_1$ and $\hat{\Sc}_2$ are given  reconstruction alphabets.
	\end{enumerate}

	
	For a given code,  let the random message $W_k$, for $k=1,2$, be uniform over the message set $\Wc_k$  and the inputs $X_{k,i}=\phi_{k,i}(W_k, Z_{k}^{i-1})$,  for $i=1, \ldots,  n$. The Txs' state estimates 
	are obtained as $\hat{S}_k^n:= (\hat{S}_{k, 1}, \cdots, \hat{S}_{k, n} )=\phi_k(X_k^n,  Z_k^n)$ and the Rx's guess of the messages as  $(\hat{W}_1, \hat{W}_2)=g(Y^n)$.
	
	We shall measure the quality of the state estimates $\hat{S}_k^n$   by  bounded per-symbol distortion functions $d_k\colon \Sc_k\times \hat{\Sc}_k \mapsto [0, \infty)$,m 
	and consider \emph{expected average block distortions}
	\begin{equation}
		\Delta_k^{(n)}:= \frac{1}{n} \sum_{i=1}^n \mathbb{E}[d_k(S_{k, i},  \hat{S}_{k, i})],  \quad k=1, 2.
	\end{equation}
	The probability of decoding error is defined as:
	\begin{IEEEeqnarray}{rCl}
		P^{(n)}_e& := &\textnormal{Pr}\Big( \hat{W}_1 \neq W_1 
		\quad \textnormal{or} 
		\quad \hat{W}_2\neq W_2 \Big).
	\end{IEEEeqnarray}
	\begin{definition} 
		\hspace{0.5cm}	A rate-distortion tuple $(\R_1,  \R_2,  \D_1,  \D_2)$ is\\
		achievable if there exists  a sequence (in $n$) of  $(2^{n\R_1}, 2^{n\R_2},  n)$ codes that simultaneously satisfy
		\begin{subequations}\label{eq:asymptotics}
			\begin{IEEEeqnarray}{rCl}
				\lim_{n\to \infty}	P^{(n)}_e 
				&=&0 \\
				\varlimsup_{n\to \infty}	\Delta_k^{(n)}& \leq& \D_k,  \quad \textnormal{for } k=1, 2.\label{eq:asymptotics_dis}
			\end{IEEEeqnarray}
		\end{subequations}
	\end{definition}
	\begin{definition}
		The capacity-distortion region $\CDc$ is the closure of the set of all achievable tuples $(\R_1,  \R_2, \D_1, \D_2)$.
	\end{definition}
	
	The main result of this paper is the inner bound on the capacity-distortion region $\CDc$ given in the following Theorem~\ref{Th:achievability}. A scheme achieving this region is described in Section~\ref{Achieve}, for the analysis see \sh{\cite[Appendix~A]{arxiv}.} \lo{Appendix~\ref{app:analysis}.} It is based on a modification of Willem's coding scheme \cite{willems1983achievable} for the MAC with generalized feedback. That means block-Markov encoding and backward decoding are
	used where each Tx splits its message into a private and a common part for each block (except for the last block). Tx~$k$ sends its common \mw{message parts} using   $U_k$-codewords and after each block decodes the common message part sent by the other Tx~$\bar{k}$ based on its generalized feedback outputs. This allows the two Txs to cooperatively send \emph{both} common parts from the previous block using the $U_0$-codewords. Private parts are sent using the $X_1$- and $X_2$-codewords and are only decoded at the Rx. 
	The novelty of our scheme with respect to \cite{willems1983achievable} is that together with its common part pertaining to the current block, each Tx~$k$ also sends a $V_k$-compression codeword containing information about the other Tx's desired state $S_{\bar{k}}$ of the previous block. This compression information is decoded at both the other Tx~$\bar{k}$ and at the Rx. \mw{At the Rx it is  used to improve  the decoding of the messages.} 
	
	%
	\begin{theorem}\label{Th:achievability}
		The capacity-distortion region $\Cc\Dc$ includes any rate-distortion tuple $(R_1, R_2, D_1, D_2)$ that for some choice of  pmfs 
		$P_{U_0}, P_{U_1\mid U_0},
		P_{U_2\mid U_0}, P_{X_1\mid U_0U_1}, P_{X_2\mid U_0U_2},$ $P_{V_1\mid U_0U_2X_1Z_1}, P_{V_2\mid U_0U_1X_2Z_2}$ and  estimators $\phi_k^*\colon \mathcal{X}_k \times \mathcal{Z}_k \times \mathcal{U}_{\bar{k}} \times  \mathcal{V}_{\bar{k}}\to \hat{\mathcal{S}}_k$,  for $k=1,2$, 
		satisfies  Inequalities \eqref{eq:inner} on top of the next page \mw{(where $\underline{U}:=(U_0,U_1,U_2)$)}
		\begin{figure*}[t]
			\mw{			%
				\begin{subequations}\label{eq:inner}
					\begin{IEEEeqnarray}{rCl}
						R_k&\leq&I(U_k;X_{\bar{k}}Z_{\bar{k}}\mid U_0U_{\bar{k}})
						+
						I(V_k;X_{\bar{k}}Z_{\bar{k}}\mid \underline{U} )
						-I(V_k;X_kZ_k\mid \underline{U})
						\nonumber\\
						&&+\min\{
						I(X_k;Y\mid U_0X_{\bar{k}})
						+ I( V_k; X_1X_2Y \mid \underline{U}) 	+ I( V_{\bar{k}}; X_1X_2YV_k \mid \underline{U})
						-I(V_k;X_kZ_k\mid \underline{U}),
						\nonumber\\
						&&\hspace{1.2cm} I(X_1X_2;Y\mid U_0U_k)
						+I( V_k; X_1X_2Y \mid \underline{U}) 	+ I( V_{\bar{k}}; X_1X_2YV_k\mid \underline{U})
						-I(V_{\bar{k}};X_{\bar{k}}Z_{\bar{k}}\mid \underline{U}), \nonumber\\
						&&
						\hspace{1.2cm}  I(X_1X_2;Y\mid U_0)+ I( V_k; X_1X_2Y \mid \underline{U}) 	+ I( V_{\bar{k}}; X_1X_2YV_k\mid \underline{U})-I(V_k;X_kZ_k\mid \underline{U}) 
						-I(V_{\bar{k}};X_{\bar{k}}Z_{\bar{k}}\mid \underline{U})\nonumber \\
						& & \hspace{1.2cm}   I(X_k;YV_1V_2\mid \underline{U}X_{\bar{k}})
						\} , \hspace{5cm}  k=1,2,\IEEEeqnarraynumspace
						\vspace{0.5cm}\\
						%
						%
						R_1+R_2&\leq& I(U_2;X_1Z_1\mid U_0U_1)
						+I(V_2;X_1Z_1\mid \underline{U})
						- I(V_2;X_2Z_2\mid \underline{U})
						\nonumber\\
						&& +I(U_1;X_2Z_2\mid U_0U_2)	+I(V_1;X_2Z_2\mid \underline{U})
						- I(V_1;X_1Z_1\mid \underline{U})
						\nonumber\\
						&&+	\min\{I(X_1X_2;Y \mid U_0U_2)
						+I( V_1; X_1X_2Y \mid \underline{U}) 	+ I( V_2; X_1X_2YV_1 \mid \underline{U})
						-I(V_1;X_1Z_1\mid \underline{U}) ,
						\nonumber\\&&
						\hspace{1.2cm}
						I(X_1X_2;Y\mid U_0U_1)
						+I( V_1; X_1X_2Y \mid \underline{U}) 	+ I( V_2; X_1X_2YV_1 \mid \underline{U})	
						-I(V_2;X_2Z_2\mid \underline{U}),
						\nonumber\\
						&&
						\hspace{1.2cm} I(X_1X_2;Y\mid U_0)
						+I( V_1; X_1X_2Y \mid \underline{U}) 	+ I( V_2; X_1X_2YV_1 \mid \underline{U})	
						-I(V_1;X_1Z_1\mid \underline{U}) -I(V_2;X_2Z_2\mid \underline{U})
						\nonumber\\
						&& \hspace{1.2cm} I(X_1X_2;YV_1V_2\mid \underline{U})
						\} \vspace{0.3cm}	\\
						R_{1}+R_{2}&\leq&
						I(X_1X_2; Y )
						+ I( V_1; X_1X_2Y \mid \underline{U}) 	
						+ I( V_2; X_1X_2YV_1 \mid \underline{U})
						-I(V_1;X_1Z_1\mid \underline{U}) -I(V_2;X_2Z_2\mid \underline{U})\label{eq:R122}\IEEEeqnarraynumspace
					\end{IEEEeqnarray}
					and
					\begin{IEEEeqnarray}{rCl}
						I(U_k;X_{\bar{k}}Z_{\bar{k}}\mid U_0U_{\bar{k}})+I(V_k;X_{\bar{k}}Z_{\bar{k}}\mid \underline{U}) &\geq& I(V_k;X_kZ_k\mid \underline{U}),  \hspace{1cm} k=1,2,\\
						I(X_1X_2;Y\mid U_0)+
						I( V_1; X_1X_2Y \mid \underline{U}) 	+ I( V_2; X_1X_2YV_1 \mid \underline{U})
						& \geq& I(V_1;X_1Z_1\mid \underline{U}) +I(V_2;X_2Z_2\mid \underline{U})\\
						I(X_k;Y\mid U_0X_{\bar{k}})
						+
						I( V_1; X_1X_2Y \mid \underline{U}) 	+ I( V_2; X_1X_2YV_1 \mid \underline{U}) &\geq& I(V_k;X_kZ_k\mid \underline{U}),  \hspace{1cm} k=1,2.
					\end{IEEEeqnarray}
				\end{subequations}
			}
			\vspace{-3mm}
			\hrule
			\vspace{-2mm}
		\end{figure*}
		as well as the distortion constraints
		\begin{eqnarray}\label{Th:distortion}
			&	 \textnormal{E}[d_k(S_k, \phi^*_{k}(X_k, Z_k, U_{\bar{k}}, V_{\bar{k}}
			)
			]\leq D_k,  \quad k=1,2,\label{eq:Dk}
		\end{eqnarray}
		where all quantities are evaluated for  
		$(U_0, U_1 ,U_2,X_1, X_2, Y,Z_1, Z_2, V_1, V_2) 
		\sim P_{U_0}P_{U_1\mid U_0} 
		P_{U_2\mid U_0}$
		$P_{X_1\mid U_0 U_1}
		P_{X_2\mid U_0 U_2}
		P_{S_1S_2}
		P_{YZ_1Z_2\mid S_1S_2X_1X_2}P_{V_1\mid U_0U_2X_1Z_1}$ $
		P_{V_2\mid U_0U_1X_2Z_2}$
		and for $k=1,2$:
		\begin{IEEEeqnarray}{rCl}\label{Th:estimator}\lefteqn{
				\phi_k^*(x_k,z_k,u_{\bar{k}},v_{\bar{k}}):= }\nonumber \\ 
			&&	\textnormal{arg}\min_{s_k'\in \hat{\mathcal{S}_k}} \sum_{s_k\in \mathcal{S}_k}  P_{S_k|X_kZ_kU_{\bar{k}} V_{\bar{k}}}(s_k|x_k,z_k,u_{\bar{k}},v_{\bar{k}})\;  d_k(s_k,  s_k').
			\nonumber \\
		\end{IEEEeqnarray}
	\end{theorem}
	\begin{remark} Our model includes as special cases all setups  with perfect or imperfect channel state-information at the receiver. 
		For example, for the choise of 
		\begin{equation}
			Y=(Y', S_1, S_2)
		\end{equation}
		with $Y'$  describing any desired output, the receiver has perfect CSI about both states.
	\end{remark}
	%
	%
	%
	%
	\begin{corollary}\label{cor}
		For $V_1=V_2=$const, Theorem~\ref{Th:achievability} specializes to \cite[Theorem~2]{kobayashi2019joint}, i.e., to
		the set of   tuples $(R_1, R_2, D_1, D_2)$ that for some  pmfs $P_{U_0}P_{U_1\mid U_0}P_{U_2\mid U_0}P_{X_1\mid U_1U_0}P_{X_2\mid U_2U_0}$
		satisfy 
		\begin{IEEEeqnarray}{rCl}
			R_k&\leq & I(X_k; Y\mid X_{\bar{k}}U_kU_0)+I(U_k;Z_{\bar{k}}\mid X_{\bar{k}}U_0), \nonumber \\
			& &\hspace{4cm} k=1,2,
			\\
			R_1+R_2&\leq& I(X_1X_2;Y),
			\\
			R_1+R_2&\leq& I(X_1X_2;Y\mid U_0U_1U_2)
			\nonumber\\&&\hspace{0cm}+I(U_1;Z_2\mid X_2U_0)
			+I(U_2;Z_1\mid X_1U_0),
		\end{IEEEeqnarray}
		and
		\begin{eqnarray}
			\textnormal{E}\Big[d_k(S_k, \phi_k^*(X_k, Z_k,U_{\bar{k}}, \textnormal{const}))\Big]\leq D_k, \quad k=1,2,
		\end{eqnarray}
		where all quantities are evaluated for $(U_0, U_1, U_2, X_1, $ $X_2,S_1, S_2, Y, Z_1, Z_2)\sim	P_{U_0}P_{U_1\mid U_0}
		P_{U_2\mid U_0}P_{X_1\mid U_1U_0}P_{X_2\mid U_2U_0}$
		$P_{S_1S_2}P_{YZ_1Z_2\mid X_1X_2S_1S_2}$.
		
	\end{corollary}

	The following two examples show the advantage of   Theorem~\ref{Th:achievability}  compared to Corollary~\ref{cor}.
	\begin{example}
		Consider a \mw{memoryless multiple-access channel} with binary input, output, and state alphabets $\Xc_1=\Xc_2=\Yc=\Sc_2=\{0, 1\}$. State $S_2\sim Ber(p_s)$, while $S_1=0$ is a constant. The channel input-output relation is described by 
		\begin{equation}\label{ex1:channel}
			Y= S_2 X_2,  
			\qquad
			(Z_1,Z_2)=(S_2,X_1).
		\end{equation} 	
		\mw{For this channel,} the following tuple
		\begin{equation}\label{eq:t1}
			(\R_1,  \R_2,  \D_1,  \D_2)= (0,  0,  0, 0),
		\end{equation}
		lies in the achievable region of Theorem~\ref{Th:achievability} but not in the region of Corollary~\ref{cor}, i.e., not in the region reported in \cite{kobayashi2019joint}. More specifically, choosing  $V_1=Z_1=S_2$ and the estimators $(\hat{S}_2=V_1, \hat{S}_1=0)$  in Theorem~\ref{Th:achievability} proves achievability of the desired quadruple. In contrast, $\D_2=0$ is not achievable in Corollary~\ref{cor} because  $S_2$ is independent  of $(U_1, U_2, U_0, X_1, X_2)$ and thus of $(X_2, U_1, Z_2)$, and the \mw{optimal estimator is the trivial estimator $\hat{S}_2=\psi_2^*( X_2,Z_2,U_1)=\mathbbm{1}\{ p_s > 1/2)$ which achieves} distortion $\D_2=\min\{1-p_s, p_s\}$.
		%
	\end{example}
	
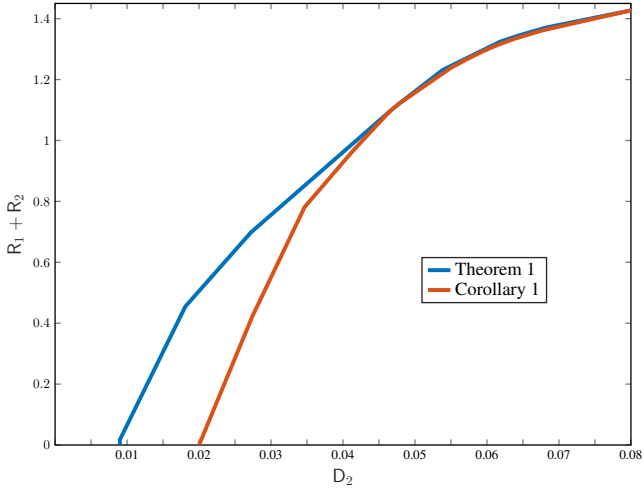
\begin{figure}[h!]
		
		%
		\definecolor{mycolor1}{rgb}{0.00000,0.44700,0.74100}%
		\definecolor{mycolor2}{rgb}{0.85000,0.32500,0.09800}%
		\centering
		
		\begin{tikzpicture}[scale=0.5]
			\pgfplotsset{every axis/.append style={
					line width=1 pt,
					tick style={line width=1pt}}}
			\begin{axis}[%
				width=6.028in,
				height=4.626in,
				at={(0in,0in)},
				scale only axis,
				xmin=0,
				xmax=0.08,
				xlabel style={font=\color{white!15!black}},
				xlabel={\Large$\D_2$},
				ymin=0,
				ymax=1.45,
				ylabel style={font=\color{white!15!black}},
				ylabel={\Large$\R_1+\R_2$},
				axis background/.style={fill=white},
				legend style={at={(0.636,0.321)}, anchor=south west, legend cell align=left, align=left, draw=white!15!black},
				every node near coord/.append style={font=\tiny},
				x tick label style={/pgf/number format/.cd,%
					scaled x ticks = false,
					set decimal separator={.},
					fixed},
				xticklabels={ , ,  ,0.01, , 0.02,   ,0.03, , 0.04,   ,0.05, , 0.06, ,0.07, , 0.08},
				tick label style={/pgf/number format/fixed},
				every node near coord/.append style={font=\large},
				]
				\addplot [color=mycolor1, line width=3.0pt]
				table[row sep=crcr]{%
					0.009 0\\
					0.009	0.0174636602105882\\
					0.009	0.0174636602105883\\
					0.009	0.0174636602105884\\
					0.0181	0.454923085608388\\
					0.0272	0.697925727944108\\
					0.046536	1.09649065861506\\
					0.04784	1.12169603783657\\
					0.05386544	1.23201509074945\\
					0.05396544	1.23301509074945\\
					0.06184672	1.32467214365213\\
					0.06484048	1.34787596202844\\
					0.06829488	1.37119420097369\\
					0.080276	1.42848686787232\\
					0.081404	1.42860128650098\\
					%
					%
					%
					%
					%
					0.1	0\\
				};
				\addlegendentry{\Large Theorem 1}
				
				\addplot [color=mycolor2, line width=3.0pt]
				table[row sep=crcr]{%
					0.02	0\\
					0.0274	0.422096034230353\\
					0.03466	0.781081613885977\\
					0.04132	0.964385233112143\\
					0.046536	1.09649065861506\\
					0.04784	1.12169603783657\\
					0.0550376	1.23920531057077\\
					0.057228	1.26697188314791\\
					0.05954	1.29381001683878\\
					0.060576	1.30468615438059\\
					0.0617024	1.31545915025154\\
					0.0638056	1.33385787412638\\
					0.0678344	1.3621799517021\\
					0.080276	1.42848686787232\\
					0.081404	1.42860128650098\\
				};
				\addlegendentry{\Large Corollary 1}
				
			\end{axis}

		\end{tikzpicture}%
		\caption{Sum-rate distortion tradeoff achieved by Theorem~\ref{Th:achievability} and Corollary~\ref{cor}, see \cite{kobayashi2019joint}, in Example~\ref{ex2} with $p_s=0.9$ and $t=0.2$ for the described choices of  auxiliaries.}	\label{fig:plot}
	\end{figure}

	We next consider the example in \cite{kobayashi2019joint}. 
		\begin{example}\label{ex2}
			Consider binary noise, states and channel inputs  $B, S_k , X_k \in \{0, 1\}$, {where $B$ is distributed Bernoulli-$t$ independent of the states and} $S_1,S_2$ are i.i.d.  Bernoulli-$p_s$, for $t, p_s\in(0,1)$. The  outputs are described as 
			{\begin{IEEEeqnarray}{C}
					Y'=S_1X_1+ S_2X_2,\\ 	Y  =  (Y', S_1,S_2),\qquad
					Z_1  = Y', \qquad 
					Z_2=Y'+B.
			\end{IEEEeqnarray}}
			%
			We again consider Hamming distortion.
			
			We further focus on binary auxiliaries $U_0,U_1, U_2$ and  $X_k=U_k\oplus\Xi_k$, for $k=1,2$ and independent binary random variables $\Xi_k$, similarly to \cite{kobayashi2019joint},\footnote{In  \cite{kobayashi2019joint} they were referred to as $U, V_1, V_2$}
			and choose the  compression variables
			\begin{equation}\label{eq:compression} V_1=\begin{cases} \mathbbm{1}\{Y'=1\} & \textnormal{ if } E=0 \\ ``\textnormal{?}" & \textnormal{ if } E=1\end{cases}
				\qquad \qquad V_2=0, 
			\end{equation} for a binary  $E$ independent of $(S_1, S_2, B, U_0, U_1, U_2,\Xi_1,\Xi_2)$.
			For this choice,  Tx 1 conveys  information about  $Y$ to  Tx~2, which  helps this latter to better estimate its state $S_2$.  In fact, when $E=0$, Tx 2 learns perfectly $Y$  because  
			\begin{equation}
				Y= \begin{cases} 0 & \textnormal{ if } Z_2\in\{0,1\}, V_1=0 \\  1 & \textnormal{ if } V_1=1\\
					2 & \textnormal{ if } Z_2\in\{2,3\}, V_1=0\end{cases}
			\end{equation} 
			

			For $p_s=0.9$ and $t=0.2$ and above choices of random variables, Figure~\ref{fig:plot} shows the maximum sum-rate $R_1+R_2$ in function of distortion $\D_2$ achieved by Theorem~\ref{Th:achievability} and Corollary~\ref{cor}, see \cite{kobayashi2019joint}.  (Corollary~\ref{cor} is simply obtained by setting $V_1=0$.)

			Notice that minimum distortion $\D_2$ in  Corollary~\ref{cor} is achieved by setting $X_1=0$ and $X_2=1$ deterministically, and is given by
			\begin{IEEEeqnarray}{rCl}\label{noV_dist}
				D_{2,\textnormal{min}}^{\textnormal{Cor}}&=& \min\left\{ p_s \bar{t}, \bar{p}_s t\right\},
			\end{IEEEeqnarray}
			which evaluates to $0.02$ for our example with $p_s=0.9$ and $t=0.2$. 
			To achieve minimum distortion $\D_2$ in Theorem~\ref{Th:achievability}, it is still optimal to choose a deterministic $X_2=1$, however, $X_1$ should not be deterministic so as to allow Tx~1 to convey information about  $Y$ to  Tx~2.  Restricting to $\Prv{E=0}=1$ and  $V_1$ in \eqref{eq:compression}, any input $X_1$ is permissible that satisfies 
			\begin{equation}\label{eq:co}
				I(V_1;Y|Z_1) \leq I(X_1;Z_2|X_2).
			\end{equation}
			The corresponding minimum distortion is given by
			\begin{IEEEeqnarray}{rCl}\label{withVdist}
				D_{2,\textnormal{min}}^{(\ref{eq:compression})}&=&  \Pr[X_1=1]\cdot p_s \bar{p}_s . \IEEEeqnarraynumspace
			\end{IEEEeqnarray}For our example, we require  {$\Pr[X_1=1]\geq 0.1 $ }for \eqref{eq:co} to hold, and the resulting minimum distortion is $D_{2,\textnormal{min}}^{(\ref{eq:compression})}=0.009$.
			

	\end{example}

	\section{Proof of Theorem~\ref{Th:achievability}}\label{Achieve}
	
	Choose a large number of blocks $B$ and split the blocklength $n$ into $B+1$ blocks of size $N:=n/(B+1)$ each. Accordingly, let $X_{1,(b)}^N, X_{2,(b)}^N, S_{1,(b)}^N, S_{2,(b)}^N, Y_{(b)}^N, Z_{1,(b)}^N, Z_{2,(b)}^N$ denote the block-$b$ inputs, states and outputs, e.g., $S_{1,(b)}^N:=(S_{1(b-1)N+1}, \ldots, S_{1, bN})$.
	
	Fix a rate-distortion tuple $(R_1,R_2,D_1,D_2)$ and pmfs  $P_{U_0}, P_{U_1\mid U_0}, P_{U_2\mid U_0},$ $P_{X_1\mid U_0U_1}, P_{X_2\mid U_0U_2}, P_{V_1\mid U_0U_2X_1Z_1}$, $P_{V_2\mid U_0U_1X_2Z_2}$ satisfying Constraints \eqref{eq:inner}  and \eqref{Th:distortion} in Theorem~\ref{Th:achievability} with strict inequality.    As shown in \lo{Appendix~\ref{app:FME}}\sh{\cite[Appendix~B]{arxiv}} using the Fourier-Motzkin Elimination algorithm, it is then possible to choose nonnegative auxiliary rates $R_{1,c}, R_{1,p}, R_{2,c}, R_{2,p}, R_{1,v}, R_{2,v}$ satisfying 
	\begin{subequations}\label{eq:subrates}
		\begin{equation}\label{eq:d1}
			R_{k,p} +R_{k,c} = R_k, \quad k=1,2,
		\end{equation} 
		and for $k=1,2$:
		\begin{IEEEeqnarray}{rCl}
			R_{k,v} &>&  I(V_k;X_kZ_k \mid \underline{U}) \label{eq:Rkv}\\
			\mw{R_{\bar{k},v}+} R_{k,c}&<& I(U_kV_{\bar{k}};X_{\bar{k}}Z_{\bar{k}}\mid U_0U_{\bar{k}})\label{eq:Rkc}\\
			%
			\mw{R_{1,v}+} R_{2,v}+R_{k,c}&<& I(U_kV_{\bar{k}}; X_{\bar{k}}Z_{\bar{k}}\mid U_0U_{\bar{k}})
			\nonumber\\
			&&	\hspace{0cm}+I(V_k;X_{\bar{k}}Z_{\bar{k}}\mid   \underline{U})\label{eq:RkvRkc}
			\\
			R_{k,p} &< & I(X_k ; YV_1V_2\mid \underline{U} X_{\bar{k}} ) \label{eq:Rkp}
			\\
			%
			%
			R_{k,v}+R_{k,p}&<& I(X_k; Y \mid U_0 X_{\bar{k}} ) 	\nonumber\\
			&&	+ I( V_2; X_1X_2YV_1   \mid \underline{U})\nonumber \\
			&&+I( V_1; X_1X_2Y \mid   \underline{U}) \IEEEeqnarraynumspace
			\\
			R_{k,v}+R_{k,p}+R_{\bar{k},p}&<&
			I(X_1X_2; Y \mid U_0 U_{\bar{k}}) 	\nonumber\\
			&&+ I( V_2 ;X_1X_2YV_1 \mid \underline{U}) \nonumber\\ 
			&&	+ I( V_1; X_1X_2Y \mid  \underline{U}) 
		\end{IEEEeqnarray}
		and 
		\begin{IEEEeqnarray}{rCl}
			\lefteqn{R_{1,p}+R_{2,p} < I(X_1 X_2; YV_1V_2 \mid \underline{U}  )}\label{eq:R1pR2p}
			\\
			\lefteqn{R_{1,v} +R_{1,p}+ R_{2,v}+R_{2,p}	}
			\nonumber \\  
			&&\qquad\qquad <	I(X_1X_2; Y \mid U_0 )  + I( V_1; X_1X_2Y \mid   \underline{U}) 
			\qquad \nonumber\\
			&&\qquad\qquad	\qquad\qquad\qquad\quad+ I( V_2; X_1X_2YV_1   \mid \underline{U}) 
			\\
			\lefteqn{R_{1,v}+R_1 +R_{2,v}+R_2}\nonumber\\
			& &\qquad\qquad <  I(X_1X_2; Y )
			+ I( V_1; X_1X_2Y \mid   \underline{U}) 
			\qquad \nonumber\\
			&&\qquad\quad	\qquad\qquad\qquad\quad
			+ I( V_2; X_1X_2YV_1 \mid \underline{U})\label{eq:R1R2R2v}\nonumber\\
			\IEEEeqnarraynumspace
		\end{IEEEeqnarray}
	\end{subequations}
	recall we used the abbreviation $\underline{U}:=(U_0,U_1,U_2)$. As we will see, Constraint~\eqref{eq:Rkv} ensures that  Tx~$k$ finds an adequate $V_k$-compression codeword. Constraints~\eqref{eq:Rkc} and \eqref{eq:RkvRkc} ensure that based on its  feedback signal and channel inputs, Tx~$k$  can decode the common message and the compression information sent by the other Tx~$\bar{k}$.  Constraints~\eqref{eq:Rkp}--\eqref{eq:R1R2R2v} ensure that the Rx can decode all transmitted messages as well as the transmitted compression informations.

	Let  each Tx $k=1,2$ split its message into $2B$ independent submessages 
	$W_k=\{(W_{k,p,(b)}, W_{k,c,(b)})\}_{b=1}^{B}$ where each $W_{k,p,(b)}$ is uniformly distributed over $[2^{NR_{k,p}}]$ and each $W_{k,c,(b)}$ is uniformly distributed over $[2^{NR_{k,c}}]$.

	Based on the conditional pmfs chosen above, define:
	\begin{IEEEeqnarray}{rCl}\label{eq:pmf}
		&&	P_{U_0U_1U_2X_1X_2S_1S_2YZ_1Z_2}:=
		\nonumber\\
		&&P_{U_0} P_{U_1\mid U_0}
		P_{U_2\mid U_0} 
		P_{X_1\mid U_0U_1} 
		P_{X_2\mid U_0U_2} P_{S_1S_2} 
		\nonumber\\&&\hspace{1cm}P_{YZ_1Z_2|S_1S_2X_1X_2} 
		P_{V_1\mid U_0U_2X_1Z_1} P_{V_2\mid U_0U_1X_2Z_2}. \IEEEeqnarraynumspace
	\end{IEEEeqnarray}
	For each block $b=1, \cdots, B+1$, do the following. 
	
	Generate an independent length-$N$ sequence $u^N_{0,(b)}(w_{1,c},w_{2,c})$ for each pair $w_{1,c}\in[ 2^{NR_{1,c}}]$ and $w_{2,c}\in [2^{NR_{2,c}}]$ by drawing each entry  i.i.d.  $P_{U_0}(\cdot )$.

	For each pair $(w_{1,c},w_{2,c})\in [2^{NR_{1,c}}]\times [2^{NR_{2,c}}]$ and each user $k=1,2$:  
	Generate a sequence $u^N_{k,(b)}(w'_{k,c}, j_{k}\mid w_{1,c},w_{2,c})$ for each pair $w'_{k,c}\in[2^{NR_{k,c}}]$ and $j_{k}\in [2^{NR_{k,v}}]$, by drawing the $i$-th entry  of this sequence according to $P_{U_k\mid U_0}(\cdot \mid u_0)$ for  $u_0$ denoting the $i$-th entry of  $u_0^N(w_{1,c},w_{2,c})$.
	
	Further, for each pair $w'_{k,c}\in[2^{NR_{k,c}}]$ and $j_{k}\in [2^{NR_{k,v}}]$ generate  
	a  sequence $x^N_{1,(b)}(w'_{1,p}\mid w'_{1,c}, j_{1}, w_{1,c},w_{2,c})$ for each index $w'_{k,p} \in [2^{ N R_{1,p}}]$, by drawing the $i$-th entry of this sequence according to $P_{X_k\mid U_0U_k}(\cdot \mid u_0,u_k)$ for $u_0$ and $u_k$ denoting the $i$-th entries of the sequences  $u^N_{0,(b)}(w_{1,c},w_{2,c})$ and  $u^N_{k,(b)}(w'_{k,c}, j_{k}\mid w_{1,c},w_{2,c})$, respectively.
	
	For each sixtuple $(w_{1,c}, w_{2,c},  w'_{1,c}, j_1 w'_{2,c}, j_2)\in[2^{NR_{1,c}}]\times [2^{NR_{2,c}}]\times [2^{NR_{1,c}}]\times[2^{NR_{1,v}}]\times  [2^{NR_{2,c}}]\times [2^{NR_{2,v}}]$  generate a  sequence $v_{1,(b)}^N(j'_{1}\mid  w'_{1,c}, j_{1}, w'_{2,c}, j_2,w_{1,c}, w_{2,c})$ for each $j_1' \in [2^{NR_{1,v}}]$ and  a  sequence $v_{2,(b)}^N(j'_{2}\mid w'_{1,c}, j_{1}, w'_{2,c}, j_2, w_{1,c}, w_{2,c})$ for each $j_2' \in [2^{NR_{2,v}}]$. The sequences $v_{1,(b)}^N(j'_{1}\mid w'_{1,c}, j_{1}, w'_{2,c}, j_2,w_{1,c}, w_{2,c})$ and  $v_{2,(b)}^N(j'_{2}\mid w'_{1,c}, j_{1}, w'_{2,c}, j_2,w_{1,c}, w_{2,c})$  are obtained by drawing  their $i$-th entries  according to $P_{V_1\mid U_0U_1 U_2}(\cdot \mid  u_0,u_1 ,u_2)$   and  $P_{V_2\mid U_0U_1 U_2}(\cdot \mid  u_0,u_1, u_2)$, respectively, for  $u_0, u_1,u_2$ denoting the $i$-th entries of the sequences  $u^N_{0,(b)}(w_{1,c},w_{2,c})$,   $u^N_{1,(b)}(w'_{1,c}, j_{1}\mid w_{1,c},w_{2,c})$, and $u^N_{2,(b)}(w'_{2,c}, j_{2}\mid w_{1,c},w_{2,c})$.
	
	Reveal the  sequences to all terminals.
	For ease of notation, define for each $k=1,2$  the indices $W_{k,p,(B+1)}=W_{k,c,(B+1)}=W_{k,c,(0)}=\hat{W}_{k,c,(0)}^{(\bar{k})}=J^*_{k,(0)}= \hat{J}^{(\bar{k})}_{k,(0)}=1$.
	%
	\subsection{Operations at Tx $1$ (Operations at Tx $2$ are analogous)}
	
	In block $b=1$, Tx~$1$ sends the codeword
	\begin{equation}
		X_{1,(1)}^N = 	x_{1,(1)}^N(W_{1,p,(1)}\mid W_{1,c,(1)}, 1, 1, 1).
	\end{equation}

	We next  describe the encoding in a given block $b\in\{2, \ldots, B+1\}$, where we assume that  the Tx has  previously produced the random indices 
	$\hat{W}_{2,c,(b-2)}^{(1)}$,  $\hat{W}_{2,c,(b-3)}^{(1)}$, ${J}_{1,(b-2)}^*$, ${J}_{1,(b-3)}^*$, and $\hat{J}_{2,(b-3)}^{(1)}$. 
	Using its  feedback  outputs from the previous two blocks $Z^N_{1,(b-1)}$ and (if $b>2$) $Z^N_{1,(b-2)}$, Tx 1 looks for a unique triple $(j_1^*,\hat{w}_2, \hat{j}_2) \in [2^{NR_{1,v}}]\times[2^{NR_{2,c}}]\times [2^{NR_{2,v}}]$ simultaneously satisfying Condition
	\eqref{typ1:enc_1} on top of  this page, \begin{figure*}[t]
		\begin{IEEEeqnarray}{rCl}\label{typ1:enc_1}
			\lefteqn{\bigg(
				u_{0,(b-1)}^N\Big(W_{1,c,(b-2)}, \hat{W}_{2,c,(b-2)}^{(1)}\Big),
				\;
				u^N_{1,(b-1)}\Big(W_{1,c,(b-1)}, J^*_{1,(b-2)}\; \Big| \;
				W_{1,c,(b-2)}, \hat{W}_{2,c,(b-2)}^{(1)}
				\Big)} \quad 
			\nonumber
			\\
			&&
			u_{2,(b-1)}^N\Big( \hat{w}_{2},\hat{j}_{2} \;
			\Big| \;W_{1,c,(b-2)}, \hat{W}_{2,c,(b-2)}^{(1)}
			\Big), \; 
			x_{1,(b-1)}^N\Big(W_{1,p,(b-1)}\;\Big| \;
			W_{1,c,(b-1)}, J^*_{1,(b-2)}, W_{1,c,(b-2)}, \hat{W}_{2,c,(b-2)}^{(1)}\Big), 
			\nonumber
			\\
			&& v_{1,(b-1)}^N\Big(j^*_{1}\;\Big|\; {J^*_{1,(b-2)}}, 
			W_{1,c,(b-1)}, \hat{w}_{2}, \hat{j}_{2} ,
			W_{1,c,(b-2)}, \hat{W}_{2,c,(b-2)}^{(1)}
			\Big), \; 
			Z^N_{1,(b-1)}
			\bigg) \in \mathcal{T}^N_{\epsilon}(P_{U_0U_1U_2X_1 V_1 Z_1})
		\end{IEEEeqnarray}
		\hrule
		\vspace{-3mm}
	\end{figure*}
	and if $b > 2$ also Condition~\eqref{typ2:enc_1} on this page,
	\begin{figure*}[t]
		
		\begin{IEEEeqnarray}{rCl}\label{typ2:enc_1}
			\lefteqn{ \bigg(u_{0,(b-2)}^N\Big(
				W_{1,c,(b-3)}, 
				\hat{W}_{2,c,(b-3)}^{(1)}\Big), \; 
				u_{1,(b-2)}^N\Big(
				W_{1,c,(b-2)},
				{J}_{1,(b-2)}^*
				\; \Big| \; 
				W_{1,c,(b-3)}, 
				\hat{W}_{2,c,(b-3)}^{(1)}
				\Big) ,}
			\nonumber	\quad	\\
			&&u_{2,(b-2)}^N\Big( \hat{W}_{2,c,(b-2)}^{(1)}, \hat{J}_{2,(b-3)}^{(1)}\; \Big| \; W_{1,c,(b-3)}, \hat{W}_{2,c,(b-3)}^{(1)}
			\Big), \; 
			x_{1,(b-2)}^N\Big(W_{1,p,(b-2)}\; \Big| \; W_{1,c,(b-2)}, J^*_{1,(b-3)}, W_{1,c,(b-3)}, \hat{W}_{2,c,(b-3)}^{(1)}\Big), 
			\nonumber		\\
			&&  v_{2,(b-2)}^N\Big(
			\hat{j}_{2} 
			\; \Big| \; 
			W_{1,c,(b-2)}, 
			J^*_{1,(b-3)}
			,
			\hat{W}_{2,c,(b-2)}^{(1)},\hat{J}_{2,(b-3)}^{(1)},W_{1,c,(b-3)}, \hat{W}_{2,c,(b-3)}^{(1)}\Big), \;  Z^N_{1,(b-2)} \bigg)
			\in \mathcal{T}^N_{\epsilon}(P_{U_0U_1U_2X_1 V_2 Z_1}),
		\end{IEEEeqnarray}
		\hrule
		\vspace{-3mm}
	\end{figure*}
	where $P_{U_0U_1U_2X_1V_1Z_1}$ and $P_{U_0U_1U_2X_1V_2Z_1}$ denote the marginals of the joint pmf in \eqref{eq:pmf}.
	If there is exactly one triple $(\hat{j}_2, \hat{w}_2, j_1^*)$ satisfying these  two conditions (or the single condition \eqref{typ1:enc_1} if $b=2$), the Tx sets $\hat{J}^{(1)}_{2,(b-2)}=\hat{j}_2$, $\hat{W}_{2,c,(b-1)}^{(1)}=\hat{w}_2$ and  $J_{1,(b-1)}^*=j_1^*$ to the corresponding indices and sends the block-$b$ channel inputs 
	\begin{IEEEeqnarray}{rCl}
		X_{1,(b)}^N &=& 	x_{1,(b)}^N\Big(W_{1,p,(b)}\;\Big|\; W_{1,c,(b)}, J^*_{1,(b-1)}, W_{1,c,(b-1)}, 
		\nonumber\\
		&&\hspace{5cm}\hat{W}_{2,c,(b-1)}^{(1)}\Big).\IEEEeqnarraynumspace
	\end{IEEEeqnarray}
	Otherwise it sets $J_{1,b}^*=-1$ and stops communication.
	
	After  the last  block of feedback signals $Z_{1,(B+1)}^N$, Tx 1 also  looks for a unique index $\hat{j}_2\in [2^{NR_{2,v}}]$ simultaneously satisfying Conditions	 \eqref{typ1:enc_1c} and \eqref{typ2:enc_1c} on  the top of next page.
	\begin{figure*}[t]
		\begin{IEEEeqnarray}{rCl}\label{typ1:enc_1c}
			\lefteqn{\bigg(
				u_{0,(B+1)}^N\Big(W_{1,c,(B)}, \hat{W}_{2,c,(B)}^{(1)}\Big),
				\;
				u^N_{1,(B+1)}\Big(1, J^*_{1,(B)} \; \Big| \;
				W_{1,c,(B)}, \hat{W}_{2,c,(B)}^{(1)}\Big), \;	u_{2,(B+1)}^N\Big(1,\hat{j}_{2} \;
				\Big| \;W_{1,c,(B)}, \hat{W}_{2,c,(B)}^{(1)}
				\Big),} \qquad 
			\nonumber
			\\
			&&
			x_{1,(B+1)}^N\Big(1\;\Big| \;
			W_{1,c,(B+1)}, J^*_{1,(B)}, W_{1,c,(B)}, \hat{W}_{2,c,(B)}^{(1)}\Big), 
			\;
			Z^N_{1,(B+1)}
			\bigg)  \in \mathcal{T}^N_{\epsilon}(P_{U_0U_1U_2X_1 Z_1})  \hspace{2cm}\IEEEeqnarraynumspace
		\end{IEEEeqnarray}
		\hrule
		\vspace{-3mm}
	\end{figure*}
	
	\begin{figure*}
		\begin{IEEEeqnarray}{rCl}\label{typ2:enc_1c}
			\lefteqn{ \bigg(u_{0,(B)}^N\Big(
				W_{1,c,(B-1)}, 
				\hat{W}_{2,c,(B-1)}^{(1)}\Big), \; 
				u_{1,(B)}^N\Big(
				W_{1,c,(B)},
				{J}_{1,(B)}^*
				\; \Big| \; 
				W_{1,c,(B-1)}, 
				\hat{W}_{2,c,(B-1)}^{(1)}
				\Big) ,}
			\nonumber	\qquad	\\
			&&u_{2,(B)}^N\Big( \hat{W}_{2,c,(B)}^{(1)}, \hat{J}_{2,(B-1)}^{(1)}\; \Big| \; W_{1,c,(B-1)}, \hat{W}_{2,c,(B-1)}^{(1)}
			\Big), \; 
			x_{1,(B)}^N\Big(W_{1,p,(B)}\; \Big| \; W_{1,c,(B)}, J^*_{1,(B-1)}, W_{1,c,(B-1)}, \hat{W}_{2,c,(B-1)}^{(1)}\Big), 
			\nonumber		\\
			&& v_{2,(B)}^N\Big(
			\hat{j}_{2}
			\; \Big| \; 
			W_{1,c,(B)}, 
			J^*_{1,(B-1)}
			,
			\hat{W}_{2,c,(B)}^{(1)},\hat{J}_{2,(B-1)}^{(1)},W_{1,c,(B-1)}, \hat{W}_{2,c,(B-1)}^{(1)}\Big), \;  Z^N_{1,(B)} \bigg)
			\in \mathcal{T}^N_{\epsilon}(P_{U_0U_1U_2X_1 V_2 Z_1}),
		\end{IEEEeqnarray}
		\vspace{-2mm}
		\hrule
		\vspace{-3mm}
	\end{figure*}
	Tx~1  produces  the state estimate $	\hat{S}_1^n=\big(\hat{S}_{1,(1)}^N, \ldots, \hat{S}_{1,(B+1)}^N\big)$ by 
	computing  the block-$b=1,\ldots, B$ estimates $\hat{S}^n_{1,(b)}$ via a component-wise application of the function $\phi_1^*$  in \eqref{Th:estimator} to the selected codewords $u_{0,(b)}^N, u_{2,(b)}^N, x_{1,(b)}^N, v_{2,(b)}^N$ 
	and setting the estimate in the last block to a dummy sequence
	$\hat{S}_{1,(B+1)}^N = s_1^N$
	for some arbitrary choice  $s_1\in \hat{\mathcal{S}}_1$. (Notice that this last block will not change the asymptotic sensing performance as $B\to \infty$.)

	\subsection{Decoding at the Rx}
	The Rx performs backward decoding. It starts by decoding the last block $B+1$, then block $B$, etc., until it finally decodes the first block $b=1$. 
	
	\sh{Due to lack of space, we only describe decoding of blocks $b=2, \ldots, B$, which are performed in decreasing order starting with block $B$, then block $B-1$, etc. Decoding of blocks $b=B+1$ and $b=1$ are similar; for details see \cite{arxiv}.}
	\lo{Decoding in block $B+1$ is as follows.
		Based on its block-$B+1$ channel  \mw{outputs $Y_{(B+1)}^N$}, the  Rx searches for a unique quadruple of indices 
		$(w_{1,c}, w_{2,c}, j_{1},j_{2}) 
		\in 
		[2^{NR_{1,c}}] \times [2^{NR_{2,c}}] \times 
		[2^{NR_{1,v}}]
		\times
		[2^{NR_{2,v}}]$  satisfying
		\begin{IEEEeqnarray}{rCl}\label{eq:RxdecodingBone}
			&&	\Big(
			u^N_{0,(B+1)}(w_{1,c}, w_{2,c} ), \; u^N_{1,(B+1)}(1, j_{1} \mid w_{1,c}, w_{2,c} ), \; 
			\nonumber\\
			& &	\quad u^N_{2,(B+1)}(1, j_{2} \mid w_{1,c}, w_{2,c} ),\;  x^N_{1,(B+1)}(1\mid 1, j_{1} , w_{1,c}, w_{2,c}), \nonumber\\
			& & \quad 
			x^N_{2,(B+1)}( 1\mid 1,j_{2}, w_{1,c}, w_{2,c}), \; 
			\nonumber\\
			& &\quad v^N_{1,(B+1)}(1\mid 1,  j_{1},
			1, j_2,
			w_{1,c},
			w_{2,c}
			), \; 
			\nonumber\\
			& &	\quad v^N_{2,(B+1)}(1\mid 
			1, j_1,
			1, j_2,
			w_{1,c},
			w_{2,c},
			), 
			\mw{Y^N_{(B+1)} \Big)}  \nonumber \\
			&& \mw{\in \Tc_{2\epsilon}(P_{U_0U_1U_2X_1X_2 Y})}
		\end{IEEEeqnarray}
		If such a unique quadruple exists, it sets $\hat{W}_{1,c,(B)}=w_{1,c}$, $\hat{W}_{2,c,(B)}=w_{2,c}$, $\hat{J}_{1,(B)}=j_1$, and $\hat{J}_{2,(B)}=j_2$. Otherwise it declares  the communication in error.
		
		Then it decodes the messages sent in each block $b\in \{2,\cdots, B\}$ in decreasing order (i.e., starting with block $B$, followed by block $B-1$, etc.).} Assume that during the decoding in the previous block $b+1$, the Rx has already produced guesses $\hat{W}_{1,c,(b)}, \hat{W}_{2,c,(b)}, \hat{J}_{1,(b)}, \hat{J}_{2,(b)}$. \mw{Based on the block-$b$ outputs 
		$Y^N_{(b)}$,} it looks for a unique sixtuple $(w_{1,p},w_{2,p},w_{1,c}, w_{2,c}, j_{1}, j_2) 
	\in [2^{NR_{1,p}}]
	\times
	[2^{NR_{2,p}}]\times 
	[2^{NR_{1,c}}] \times [2^{NR_{2,c}}] \times 
	[2^{NR_{1,v}}]
	\times
	[2^{NR_{2,v}}]
	$  satisfying
	\begin{IEEEeqnarray}{rCl}\label{typ:dec_b}
		\lefteqn{
			\bigg(
			u^N_{0,b}(w_{1,c}, w_{2,c}),\;
			u^N_{1,(b)}\Big(\hat{W}_{1,c,(b)},j_{1} \; \Big | \;  w_{1,c}, w_{2,c} \Big),\;
		} \;
		\nonumber \\
		& &  \quad
		u^N_{2,(b)}\Big(\hat{W}_{2,c,(b)}, j_{2} \; \Big | \;  w_{1,c}, w_{2,c}\Big),
		\nonumber \\
		& & \quad x^N_{1,(b)}\Big(w_{1,p} \; \Big | \;  \hat{W}_{1,c,(b)}, j_{1}, w_{1,c}, w_{2,c}\Big), \; 
		\nonumber \\
		& &\quad x^N_{2,(b)}\Big( w_{2,p} \; \Big | \;  \hat{W}_{2,c,(b)}, j_{2}, w_{1,c}, w_{2,c}\Big),
		\nonumber		\\
		&&\quad
		v_{1,(b)}^N\left(\hat{J}_{1,(b)} \; \Big | \;  
		\hat{W}_{1,c,(b)}, j_1,  \hat{W}_{2,c,(b)}, j_2,
		w_{1,c}, w_{2,c}
		\right), 
		\nonumber		\\
		&&\quad
		v^N_{2,(b)}\left(\hat{J}_{2,(b)} \; \Big | \;  \hat{W}_{1,c,(b)}, j_1,  \hat{W}_{2,c,(b)}, j_2,w_{1,c}, w_{2,c} 
		\right), \;
		\mw{Y^N_{(b)} \bigg) }\nonumber \\
		& &   \mw{ \in \Tc_{2\epsilon}(P_{U_0U_1U_2X_1X_2 Y}).}
	\end{IEEEeqnarray}
	If such a  unique sixtuple exists, it sets $\hat{W}_{1,c,(b-1)}=w_{1,c}$, $\hat{W}_{1,p,(b)}=w_{1,p}$, $\hat{W}_{2,c,(b-1)}=w_{2,c}$, $\hat{W}_{2,p,(b)}=w_{2,p}$, $\hat{J}_{1,(b-1)}=j_1$, and $\hat{J}_{2,(b-1)}=j_2$. Otherwise it declares  the communication in error.
	
	\lo{For the first block $b=1$, the Rx looks for a unique pair 	
		$(w_{1,p},w_{2,p}) 
		\in 
		[2^{NR_{1,c}}] \times [2^{NR_{2,c}}]$ satisfying
		\begin{IEEEeqnarray}{rCl}\label{typ:dec_bc}
			\lefteqn{
				\bigg(
				u^N_{0,(1)}( 1_{[2]}),\;
				u^N_{1,(1)}\Big(\hat{W}_{1,c,(1)},1 \; \Big | \;  1_{[2]} \Big),} \quad \nonumber \\
			&& \quad u^N_{2,(1)}\Big(\hat{W}_{2,c,(1)}, 1 \; \Big | \;  1_{[2]} \Big), \nonumber \\
			& &  \quad x^N_{1,(1)}\Big(w_{1,p} \; \Big | \;  \hat{W}_{1,c,(1)}, 1_{[3]} \Big), \; 
			x^N_{2,(1)}\Big( w_{2,p} \; \Big | \;  \hat{W}_{2,c,(1)},1_{[3]}\Big)
			\nonumber		\\
			&&\quad 
			v_{1,(1)}^N\Big(\hat{J}_{1,(1)} \; \Big | \;  
			\hat{W}_{1,c,(1)},  1,\hat{W}_{2,c,(1)},
			1_{[3]}
			\Big), 
			\nonumber		\\
			&&\quad
			v^N_{2,(b)}\left(\hat{J}_{2,(b)} \; \Big | \;  \hat{W}_{1,c,(b)},  1, \hat{W}_{2,c,(b)},1_{[3]}
			\right), Y^N_{(b)}\bigg)  \nonumber \\
			& &  \in \Tc_{2\epsilon}(P_{U_0U_1U_2X_1X_2Y}).
		\end{IEEEeqnarray}
		If such a  unique pair exists, it sets $\hat{W}_{1,p,(1)}=w_{1,p}$,  and $\hat{W}_{2,p,(1)}=w_{2,p}$. Otherwise it declares  a communication  error.}

	The Rx finally declares the messages $\hat{W}_1$ and $\hat{W}_2$ that correspond to the produced guesses $\{(\hat{W}_{k,p,(b)}, \hat{W}_{k,c,(b)})\}$.
	
	Notice that the rate of communications of our scheme are only $\frac{B}{B+1}R_1$ and $\frac{B}{B+1}R_2$, which however approach $R_1$ and $R_2$ when $B\to \infty$.
	
	\section{Summary}
	
	We proposed the first information-theoretic fully-integrated sensing and communication scheme where coding at a transmitter is not only used for data transmission but also to improve sensing (state-estimation) at the other transmitter. 
	At the hand of examples, we show the improved performances of the new scheme compared to state of the art. 
	
	\section*{Acknowledgement}
	This work has been supported by the European Research Council (ERC) under the European Union's Horizon 2020 under grant agreement No 715111 and by the  DFG under grant 
	agreement number KR 3517/11-1. 
	\appendices
	\section{Analysis of  Error Probability and State Estimation}\label{app:analysis}
	To derive an upper bound  on the average error probability (averaged over the random code construction and the state and channel realizations), we enlarge the error event to the event that for some $k=1,2$ and $b=1,\ldots, B$:
	\begin{IEEEeqnarray}{rCl}
		\hat{W}_{k,c,(b)} \neq W_{k,c,(b)}\quad & \text{or}& \quad 
		\hat{W}_{k,p,(b)}\neq W_{k,p,(b)}\quad  \nonumber \\	
		&\text{or}& \quad 	\hat{W}_{k,c,(b)}^{(\bar k)} \neq W_{k,c,(b)}	
	\end{IEEEeqnarray}
	or 
	\begin{IEEEeqnarray}{rCl}
		J_{k,(b)}^*= -1 \quad \text{or} \quad 	\hat{J}_{k,(b)}\neq J^*_{k,(b)} \quad \text{or} \quad \hat{J}_{k,(b)}^{(\bar{k})} \neq J^*_{k,(b)}.\IEEEeqnarraynumspace
	\end{IEEEeqnarray}
	For ease of notation, we define the block-$b$ Tx-error events  for $k=1,2$ and $b=1,\ldots, B$:
	\begin{IEEEeqnarray}{rCl}
		\mathcal{E}_{\textnormal{Tx},k,(b)}&&:=\Big \{  \hat{W}^{(k)}_{\bar{k},c,(b)}\neq W_{\bar{k},c,(b)} 	\;\text{ or }\;
		\hat{J}^{(k)}_{\bar{k},(b-1)}\neq J^*_{\bar{k},(b-1)} 
		\nonumber	\\
		&&\hspace{4.7cm}	\;\text{ or }\;J_{k,b}^*= -1 \Big\},\IEEEeqnarraynumspace
	\end{IEEEeqnarray}
	and
	\begin{IEEEeqnarray}{rCl}
		\mathcal{E}_{\textnormal{Tx},k,(B+1)}:=\left\{ 
		\hat{J}^{(k)}_{\bar{k},(B)}\neq J^*_{\bar{k},(B)}\right \}, \qquad k \in\{1,2\}.
	\end{IEEEeqnarray}
	
	Define also the Rx-error events for $k=1,2$ and block $b=1,\ldots, B+1$:
	\begin{IEEEeqnarray}{rCl}
		\mathcal{E}_{\textnormal{Rx},(b)}&&:=\Big\{  \hat{W}_{k,c,(b-1)}\neq W_{k,c,(b-1)} 	\;\text{ or }\; \hat{W}_{k,p,(b)}\neq W_{k,p,(b)} 	\nonumber	\\
		&&\hspace{1.6cm}\;\text{ or }\; 
		\hat{J}_{k,(b-1)}\neq J^*_{k,(b-1)}\colon \quad k=1,2\Big\}.\IEEEeqnarraynumspace
	\end{IEEEeqnarray}
	
	By the union bound and basic probability, we  find:
	\begin{IEEEeqnarray}{rCl}\label{error_P_def}
		\lefteqn{\Pr\left( \hat{W}_1 \neq W_1 \;\text{ or }\; \hat{W}_2 \neq W_2\right)} \nonumber \qquad \\
		&\leq & \sum_{b=1}^{B+1}
		\Pr\left( \mathcal{E}_{\textnormal{Tx},1,(b)}\Bigg| \bigcup_{b'=1}^{b-1} \left\{ \bar{ \mathcal{E}}_{\textnormal{Tx},1,(b')}, \; \bar{ \mathcal{E}}_{\textnormal{Tx},2,(b')} \right\}
		\right) 
		\nonumber\\
		&&	\hspace{0cm}	+
		\sum_{b=1}^{B+1} 	\Pr\left( \mathcal{E}_{\textnormal{Tx},2,(b)}\Bigg| \bigcup_{b'=1}^{b-1} \left\{  \bar{ \mathcal{E}}_{\textnormal{Tx},1,(b')}, \; \bar{ \mathcal{E}}_{\textnormal{Tx},2,(b')} \right\}
		\right) 
		\nonumber\\
		&
		&+\sum_{b=1}^{B+1} 	\Pr\left( \mathcal{E}_{\textnormal{Rx},(b)} \Bigg|   \bigcup_{b'=1}^{B+1}\left\{ \bar{ \mathcal{E}}_{\textnormal{Tx},1,(b')}, \; \bar{ \mathcal{E}}_{\textnormal{Tx},2,(b')}\right\} \right).\IEEEeqnarraynumspace
		\label{analysis:part1}
	\end{IEEEeqnarray}

	
	We   analyze the three sums separately. The first sum  is related to Tx~1's error event, the second sum  to Tx~2's error event, and the third sum to the Rx's error event.
	
	\subsubsection{Analysis of Tx~1's error event} To simplify notations, we define for each block $b\in\{2,\ldots, B+1\}$ and each triple of indices $(j_1^*, \hat{w}_2, \hat{j}_2)$ 
	the event $\mathcal{F}_{\textnormal{Tx1},(b)}(j_1^*, \hat{w}_2, \hat{j}_2)$ that the following two conditions  \eqref{typ1:enc_1b} and \eqref{typ2:enc_1b}  (only Condition  \eqref{typ1:enc_1b} for \mw{$b=1$}) hold: 
	\begin{IEEEeqnarray}{rCl}\label{typ1:enc_1b}
		&&	 \bigg(
		u_{0,(b)}^N\Big(W_{1,c,(b-1)}, \hat{W}_{2,c,(b-1)}^{(1)}\Big),
		\;
		\nonumber
		\\
		&&	\quad u^N_{1,(b)}\Big(W_{1,c,(b)}, J^*_{1,(b-1)}\; \Big| \;
		W_{1,c,(b-1)}, {W}_{2,c,(b-1)}
		\Big)  \qquad 
		\nonumber
		\\
		&&\quad
		u_{2,(b)}^N\Big( \hat{w}_{2},\hat{j}_{2} \;
		\Big| \;W_{1,c,(b-1)}, {W}_{2,c,(b-1)}
		\Big), \; 
		\nonumber
		\\
		&&	\quad x_{1,(b)}^N\Big(W_{1,p,(b)}\;\Big| \;
		W_{1,c,(b)}, J^*_{1,(b-1)}, \nonumber \\ 
		& & \hspace{5cm} W_{1,c,(b-1)}, {W}_{2,c,(b-1)}\Big), 
		\nonumber
		\\
		&& \quad
		v_{1,(b)}^N\Big(j^*_{1}\;\Big|\; {J^*_{1,(b-1)}}, 
		W_{1,c,(b)}, \hat{w}_{2}, \hat{j}_{2} ,\nonumber \\ 
		& & \hspace{5cm}
		W_{1,c,(b-1)}, {W}_{2,c,(b-1)}
		\Big), \; 
		\nonumber \\
		& &	\quad Z^N_{1,(b)}
		\bigg) 
		\in \mathcal{T}^N_{\epsilon}(P_{U_0U_1U_2X_1 V_1 Z_1})
	\end{IEEEeqnarray}
	and if $b >1$
	\begin{IEEEeqnarray}{rCl}\label{typ2:enc_1b}
		&&\bigg(
		u_{0,(b-1)}^N\Big(
		W_{1,c,(b-2)}, 
		{W}_{2,c,(b-2})\Big), \; 
		\nonumber	\\
		&& \quad u_{1,(b-1)}^N\Big(
		W_{1,c,(b-1)},
		{J}_{1,(b-1)}^*
		\; \Big| \; 
		W_{1,c,(b-2)}, 
		{W}_{2,c,(b-2)}
		\Big) 
		\nonumber	\\
		&&\quad u_{2,(b-1)}^N\Big( {W}_{2,c,(b-1)}, {J}_{2,(b-2)}
		\Big| \; W_{1,c,(b-2)}, {W}_{2,c,(b-2)}
		\Big), \; 
		\nonumber	\\
		&&	\quad x_{1,(b-1)}^N\Big(W_{1,p,(b-1)}\; \Big| \; W_{1,c,(b-1)}, J^*_{1,(b-2)}, 
		\nonumber		\\
		&&\hspace{5cm}
		W_{1,c,(b-2)}, {W}_{2,c,(b-2)}\Big), 
		\nonumber		\\
		&&\quad v_{2,(b-1)}^N\Big(
		\hat{j}_{2}
		\; \Big| \;			
		W_{1,c,(b-1)}, 
		J^*_{1,(b-2)}
		,
		{W}_{2,c,(b-1)},{J}_{2,(b-2)}^*,
		\nonumber	\\
		&&\hspace{5cm}	W_{1,c,(b-2)}, {W}_{2,c,(b-2)}\Big), \;  
		\nonumber	\\
		&&\quad 	Z^N_{1,(b-1)} \bigg)
		\in \mathcal{T}^N_{\epsilon}(P_{U_0U_1U_2X_1 V_2 Z_1}).
	\end{IEEEeqnarray}
	Notice that compared to \eqref{typ1:enc_1} and \eqref{typ2:enc_1}, here we replaced the triple $(\hat{W}_{2,c,(b-2)}^{(1)},\hat{W}_{2,c,(b-1)}^{(1)}, \hat{J}_{2,(b-2)}^{(1)})$ by their correct values $W_{2,c,(b-2)}, W_{2,c,(b-1)}, J^*_{2,(b-2)})$.
	Similarly, define the event $\mathcal{F}_{\textnormal{Tx1},(B+1)}(\hat{j}_2)$ as the event that the following two conditions are  satisfied: 
	\begin{IEEEeqnarray}{rCl}\label{typ1:enc_1d}
		&&	\bigg(
		u_{0,(B+1)}^N\Big(W_{1,c,(B)}, {W}_{2,c,(B)}\Big),
		\;
		\nonumber		\\
		&&\quad 	u^N_{1,(B+1)}\Big(1, J^*_{1,(B)} \; \Big| \;
		W_{1,c,(B)}, {W}_{2,c,(B)}\Big) \qquad 
		\nonumber
		\\
		&&\quad 
		u_{2,(B+1)}^N\Big(1,\hat{j}_{2} \;
		\Big| \;W_{1,c,(B)}, {W}_{2,c,(B)}
		\Big), \; 
		\nonumber		\\
		&&\quad 
		x_{1,(B+1)}^N\Big(1\;\Big| \;
		W_{1,c,(B+1)}, J^*_{1,(B)}, W_{1,c,(B)}, {W}_{2,c,(B)}\Big), 
		\;
		\nonumber \\
		& & 	\quad Z^N_{1,(B+1)}
		\bigg) \in \mathcal{T}^N_{\epsilon}(P_{U_0U_1U_2X_1 Z_1})
	\end{IEEEeqnarray}
	and
	\begin{IEEEeqnarray}{rCl}\label{typ2:enc_1d}
		\lefteqn{ \bigg(u_{0,(B)}^N\Big(
			W_{1,c,(B-1)}, 
			{W}_{2,c,(B-1)} \Big), \; } \quad
		\nonumber \\
		& & 	u_{1,(B)}^N\Big(
		W_{1,c,(B)},
		{J}_{1,(B)}^*
		\; \Big| \; 
		W_{1,c,(B-1)}, 
		{W}_{2,c,(B-1)}
		\Big) ,
		\nonumber	\qquad	\\
		&&u_{2,(B)}^N\Big( {W}_{2,c,(B)}^{(1)}, {J}_{2,(B-1)}^{(1)}\; \Big| \; W_{1,c,(B-1)}, \hat{W}_{2,c,(B-1)}^{(1)}
		\Big), \; 
		\nonumber \\
		& & x_{1,(B)}^N\Big(W_{1,p,(B)}\; \Big| \; W_{1,c,(B)}, J^*_{1,(B-1)}, 	\nonumber \\
		& &\hspace{4cm}W_{1,c,(B-1)}, \hat{W}_{2,c,(B-1)}^{(1)}\Big), 
		\nonumber		\\
		&&\hspace{0cm} 
		v_{2,(B)}^N\Big(
		\hat{j}_{2}
		\; \Big| \; 
		W_{1,c,(B)}, 
		J^*_{1,(B-1)}
		,
		\hat{W}_{2,c,(B)}^{(1)},{J}_{2,(B-1)}^{(1)},
		\nonumber \\
		& & \hspace{4cm}W_{1,c,(B-1)}, {W}_{2,c,(B-1)}\Big), \; 
		\nonumber\\
		&&Z^N_{1,(B)} \bigg)
		\in \mathcal{T}^N_{\epsilon}(P_{U_0U_1U_2X_1 V_2 Z_1})
	\end{IEEEeqnarray}

	We continue by noticing that  event $\bigcup_{b'=1}^{b-1} \left\{  \bar{\mathcal{E}}_{\textnormal{Tx},1,(b')}, \; \bar{\mathcal{E}}_{\textnormal{Tx},2,(b')} \right\}$ implies  that for all $b'=1, \ldots, b-1$, $k =1,2$:
	\begin{IEEEeqnarray}{rCl}
		\hat{W}^{(k)}_{\bar{k},c,(b')}& =&W_{\bar{k},c,(b')}\\
		J_{k,(b')}^*&\neq&- 1\\
		\hat{J}^{(\bar{k})}_{k,(b'-1)}&=& J_{{k},(b'-1)}^*.
	\end{IEEEeqnarray}
	Moreover, for any block $b=1,\ldots, B+1$,  event  $ \bar{\mathcal{E}}_{\textnormal{Tx},1,(b)}$ is implied by the  event that $\mathcal{F}_{\textnormal{Tx1},(b)}(j_1^*, \hat{w}_{2},\hat{j}_2)$ \emph{is  not} satisfied for any tuple $(j_1^*, \hat{w}_{2},\hat{j}_2)$ with  $(\hat{w}_{2},\hat{j}_2)=( {W}_{2,c,(b)}, J_{2,(b-1)}^*)$ or it \emph{is} satisfied for some triple $(j_1^*, \hat{w}_{2},\hat{j}_2)$ with $( \hat{w}_{2},\hat{j}_2)\neq ({W}_{2,c,(b)}, J_{2,(b-1)}^*)$. 
	Thus,  the sequence of inequalities on top of the next page holds,
	\begin{figure*}
		\begin{subequations}
			\begin{IEEEeqnarray}{rCl}
				\lefteqn{\Pr\left( \mathcal{E}_{\textnormal{Tx},1,(b)} \; \Bigg|  \; \bigcup_{b'=1}^{b-1} \left\{ \bar{ \mathcal{E}}_{\textnormal{Tx},1,(b')}, \; \bar{ \mathcal{E}}_{\textnormal{Tx},2,(b')} \right\}
					\right) }  \nonumber \\
				& = &\Pr\Bigg( \Bigg(\bigcap_{j_1^*\in [2^{nR_{v,1}}] }\bar{\mathcal{F}}_{\textnormal{Tx1},(b)}(j_1^*, W_{2,c,(b)},J^*_{2,(b-1)})  \Bigg) \nonumber \\
				&&	 \qquad  \cup  \Bigg( \bigcup_{\substack{ (j_1^*, \hat{w}_{2},\hat{j}_2) \colon \\ ( \hat{w}_{2},\hat{j}_2)\neq ({W}_{2,c,(b)}, J_{2,(b-1)}^*) }}  \mathcal{F}_{\textnormal{Tx1},(b)}(j_1^*, \hat{w}_{2},\hat{j}_2)    \Bigg)\; \Bigg| \; \bigcup_{b'=1}^{b-1} \left\{ \bar{ \mathcal{E}}_{\textnormal{Tx},1,(b')}, \; \bar{ \mathcal{E}}_{\textnormal{Tx},2,(b')} \right\}
				\Bigg)  \nonumber \\\\
				& \leq  & \Pr\left( \bigcap_{j_1^*\in [2^{nR_{v,1}}] }\bar{\mathcal{F}}_{\textnormal{Tx1},(b)}(j_1^*, W_{2,c,(b)},J^*_{2,(b-1)})  \; \Bigg| \;\bigcup_{b'=1}^{b-1} \left\{ \bar{ \mathcal{E}}_{\textnormal{Tx},1,(b')}, \; \bar{ \mathcal{E}}_{\textnormal{Tx},2,(b')} \right\}
				\right) 
				\nonumber\\
				& & + \Pr\Bigg(  \bigcup_{\substack{ (j_1^*, \hat{w}_{2},\hat{j}_2) \colon \\ ( \hat{w}_{2},\hat{j}_2)\neq ({W}_{2,c,(b)}, J_{2,(b-1)}^*) }}  \mathcal{F}_{\textnormal{Tx1},(b)}(j_1^*, \hat{w}_{2},\hat{j}_2)  \; \Bigg| \; \bigcup_{b'=1}^{b-1} \left\{ \bar{ \mathcal{E}}_{\textnormal{Tx},1,(b')}, \; \bar{ \mathcal{E}}_{\textnormal{Tx},2,(b')} \right\}
				\Bigg) \\
				& \leq & 	 \Pr\left( \bigcap_{j_1^*\in [2^{nR_{v,1}}] }\bar{\mathcal{F}}_{\textnormal{Tx1},(b)}(j_1^*, W_{2,c,(b)},J^*_{2,(b-1)})  \; \Bigg| \;\bigcup_{b'=1}^{b-1} \left\{ \bar{ \mathcal{E}}_{\textnormal{Tx},1,(b')}, \; \bar{ \mathcal{E}}_{\textnormal{Tx},2,(b')} \right\}
				\right) 
				\nonumber\\
				& & + \sum_{\substack{ (j_1^*, \hat{w}_{2},\hat{j}_2) \colon \\ \hat{w}_{2} \neq W_{2,c,(b)}, \\ \hat{j}_2 \neq  J_{2,(b-1)}^*}} 
				\Pr\left(  \mathcal{F}_{\textnormal{Tx1},(b)}(j_1^*, \hat{w}_{2},\hat{j}_2)  \; \Bigg| \; \bigcup_{b'=1}^{b-1} \left\{ \bar{ \mathcal{E}}_{\textnormal{Tx},1,(b')}, \; \bar{ \mathcal{E}}_{\textnormal{Tx},2,(b')} \right\} 
				\right) 
				\nonumber\\
				& & +  \sum_{\substack{ (j_1^*,\hat{j}_2) \colon \\  \hat{j}_2 \neq  J_{2,(b-1)}^*}} 
				\Pr\left(  \mathcal{F}_{\textnormal{Tx1},(b)}(j_1^*,\mw{W_{2,c,(b)}},\hat{j}_2)  \; \Bigg| \; \bigcup_{b'=1}^{b-1} \left\{ \bar{ \mathcal{E}}_{\textnormal{Tx},1,(b')}, \; \bar{ \mathcal{E}}_{\textnormal{Tx},2,(b')} \right\} 
				\right) 	
				\nonumber\\
				& & +  \sum_{\substack{ (j_1^*, \hat{w}_{2}) \colon \\  \hat{w}_{2} \neq W_{2,c,(b)} }} 
				\Pr\left(  \mathcal{F}_{\textnormal{Tx1},(b)}(j_1^*, \hat{w}_{2},\mw{{J}_{2,(b-1)}^*})  \; \Bigg| \; \bigcup_{b'=1}^{b-1} \left\{ \bar{ \mathcal{E}}_{\textnormal{Tx},1,(b')}, \; \bar{ \mathcal{E}}_{\textnormal{Tx},2,(b')} \right\} 
				\right) ,
				\label{eq:last}
			\end{IEEEeqnarray}
		\end{subequations}
		\hrule
	\end{figure*}
	where the inequalities hold by the union bound.
	By the Covering Lemma \cite{ElGamal}, the way we construct the codebooks and  the weak law of large numbers, and because we condition on event  $\bar{\mathcal{E}}_{\textnormal{Tx},2,(b-1)}$ implying $J_{2,b-1}^*\neq -1$, the first summand in \eqref{eq:last} tends to 0 as $N\to \infty$ if 
	\begin{equation}\label{eq:R1v}
		R_{1,v}> I(V_1;X_1Z_1\mid U_0U_1U_2).
	\end{equation}
	By the way we constructed the codebooks, and standard information-theoretic arguments \cite{ElGamal}, the  sum in the second  line of \eqref{eq:last} tends to 0 as $N \to \infty$, if 
	\begin{IEEEeqnarray}{rCl}\label{eq:R2cv}
		\mw{R_{1,v}+}	R_{2,v}+R_{2,c}&<& I(U_2V_1;Z_1X_1\mid U_0U_1)
		\nonumber\\
		&&	\hspace{1cm}+I(V_2;Z_1X_1\mid U_0U_1U_2), \IEEEeqnarraynumspace
	\end{IEEEeqnarray}	
	the sum in the third line of \eqref{eq:last} tends to 0 as $N\to \infty$ if 
	\begin{IEEEeqnarray}{rCl}\label{eq:R2vlower}
		\mw{R_{1,v}+}R_{2,v} &<& I(U_2V_1;Z_1X_1\mid U_0U_1)
		\nonumber\\
		&&	\hspace{1cm}
		+I(V_2;Z_1X_1\mid U_0U_1U_2),
	\end{IEEEeqnarray}	
	and	the sum in the fourth line of \eqref{eq:last}  tends to $0$ as $N\to \infty$ if
	\begin{IEEEeqnarray}{rCl}\label{eq:R2c}
		\mw{R_{1,v}+}	R_{2,c}&<& I(Z_1X_1;U_2V_1\mid U_0U_1).
	\end{IEEEeqnarray}
	Since Condition \eqref{eq:R2vlower} is obsolete in view of \eqref{eq:R2cv}, we conclude that for any finite $B$ the sum of the  probability of errors $\sum_{b=1}^{B+1} 
	\Pr\left( \mathcal{E}_{\textnormal{Tx},1,(b)}\big| \bigcup_{b'=1}^{b-1} \left\{ \bar{ \mathcal{E}}_{\textnormal{Tx},1,(b')}, \; \bar{ \mathcal{E}}_{\textnormal{Tx},2,(b')} \right\}
	\right)$ tends to $0$ as $N\to \infty$ if  Conditions  \eqref{eq:R1v}, \eqref{eq:R2cv}, and \eqref{eq:R2c} are satisfied.	 
	
	\subsubsection{Analysis of Tx~2's error event} 
	By similar arguments, one can also prove that for finite $B$ the sum of the  probability of errors $\sum_{b=1}^{B+1}
	\Pr\left( \mathcal{E}_{\textnormal{Tx},2,(b)}\big| \bigcup_{b'=1}^{b-1} \left\{ \bar{ \mathcal{E}}_{\textnormal{Tx},1,(b')}, \; \bar{ \mathcal{E}}_{\textnormal{Tx},2,(b')} \right\}
	\right)$ tends to $0$ as $N\to \infty$ if  Conditions  \eqref{eq:Rkv},   \eqref{eq:Rkc}, and \eqref{eq:RkvRkc}, are satisfied for $k=2$.
	
	\subsubsection{Analysis of Rx's error event} 
	Define the following events. For each quadruple $(w_{1,c},w_{2,c},j_1,j_2)\in[2^{NR_{1,c}}] \times [2^{NR_{1,c}}]  \times [2^{NR_{1,v}}]  \times [2^{NR_{2,v}}]$ define   $\mathcal{F}_{\textnormal{Rx},(B+1)}(w_{1,c}, w_{2,c},  j_1, j_2)$ as the event that Condition~\eqref{eq:RxdecodingBone} is satisifed; for each pair $(w_{1,p}, w_{2,p})$ define $\mathcal{F}_{\textnormal{Rx},(1)}(w_{1,p}, w_{2,p})$  as the event that 
	\eqref{typ:dec_bc} is satisfied but where $\hat{W}_{1,c,(b)}$ and $\hat{W}_{2,c,(b)}$ should be replaced by their correct values ${W}_{1,c,(b)}$ and ${W}_{2,c,(b)}$; finally, 		
	for each block $b=2,\ldots, B$ and each tuple $(w_{1,c}, w_{2,c}, w_{1,p}, w_{2,p}, j_1, j_2)$ define  $\mathcal{F}_{\textnormal{Rx},(b)}(w_{1,c}, w_{2,c}, w_{1,p}, w_{2,p}, j_1, j_2)$  as the event 
	\begin{IEEEeqnarray}{rCl}\label{typ:dec_bb}
		&&	\Bigg(
		u^N_{0,(b)}(w_{1,c}, w_{2,c}),\;
		u^N_{1,(b)}\Big({W}_{1,c,(b)},j_{1}\; \Big | \; w_{1,c}, w_{2,c} \Big),\;
		\nonumber \\
		& &	\quad u^N_{2,(b)}\Big({W}_{2,c,(b)}, j_{2}\; \Big | \; w_{1,c}, w_{2,c}\Big), \qquad \nonumber \\
		& & \quad  x^N_{1,(b)}\Big(w_{1,p}\; \Big  | \; {W}_{1,c,(b)}, j_{1}, w_{1,c}, w_{2,c}\Big), \; 
		\nonumber \\
		& &\quad x^N_{2,(b)}\Big( w_{2,p}\; \Big | \; {W}_{2,c,(b)}, j_{2}, w_{1,c}, w_{2,c}\Big)
		\nonumber		\\
		&&\quad
		v_{1,(b)}^N\Big({J}_{1,(b)}\; \Big | \; 
		{W}_{1,c,(b)},  {W}_{2,c,(b)},
		w_{1,c},  j_1, w_{2,c}, j_2
		\Big), 
		\nonumber		\\
		&&
		\quad 	v^N_{2,(b)}({J}_{2,(b)}\mid {W}_{1,c,(b)},  {W}_{2,c,(b)},w_{1,c}, j_1, w_{2,c} , j_2
		),\mw{ Y^N_{(b)}} \Bigg) \nonumber \\
		&& 
		\mw{ \in \Tc_{2\epsilon}(P_{U_0U_1U_2X_1X_2 Y})}.
	\end{IEEEeqnarray}
	We continue by noticing that for $b=2, \ldots, B$ event  $\bar{\mathcal{E}}_{\textnormal{Rx},(b)}$ is equivalent to the event that $\mathcal{F}_{\textnormal{Rx},(b)}(w_{1,c}, w_{2,c}, w_{1,p}, w_{2,p}, j_1, j_2)$ \emph{is  not} satisfied for the tuple 
	$(w_{1,c}, w_{2,c}, w_{1,p}, w_{2,p}, j_1, j_2)
	=(W_{1,c,(b-1)}, 
	{W}_{2,c,(b-1)}, $
	$W_{1,p,(b)}, W_{2,p,(b)}, J_{1,(b-1)}^*, J_{2,(b-1)}^*)$ or it \emph{is} satisfied for some tuple  $(w_{1,c}, w_{2,c}, w_{1,p}, w_{2,p}, j_1, j_2) \neq 
	(W_{1,c,(b-1)}, {W}_{2,c,(b-1)}, $
	$W_{1,p,(b)}, W_{2,p,(b)}, J_{1,(b-1)}^*, J_{2,(b-1)}^*)$.  
	Similarly for events $\bar{\mathcal{E}}_{\textnormal{Rx},(1)}$  and $\bar{\mathcal{E}}_{\textnormal{Rx},(B+1)}$.
	Thus, for $b\in\{2,\ldots, B\}$, the sequence of (in)equalities \eqref{event_error_rec}  holds,
	\begin{figure*}[t]
		\begin{subequations}\label{event_error_rec}
			\begin{IEEEeqnarray}{rCl}
				\lefteqn{\Pr\left( \mathcal{E}_{\textnormal{Rx},(b)} \; \Bigg|  \; \bigcup_{b'=1}^{B+1} \left\{ \bar{ \mathcal{E}}_{\textnormal{Tx},1,(b')}, \; \bar{ \mathcal{E}}_{\textnormal{Tx},2,(b')} \right\}
					\right) }  \nonumber \\
				& = &\Pr \Bigg(  \Bigg( \bigcup_{\substack{ (w_{1,c}, w_{2,c}, w_{1,p}, w_{2,p}, j_1, j_2)   \neq \\(W_{1,c,(b-1)}, {W}_{2,c,(b-1)}, W_{1,p,(b)}, W_{2,p,(b)}, J_{1,b-1}^*, J_{2,(b-1)}^*)  }} \mathcal{F}_{\textnormal{Rx},(b)}(w_{1,c}, w_{2,c}, w_{1,p}, w_{2,p}, j_1, j_2)   \Bigg)\nonumber \\
				& & \hspace{1cm} \cup \quad \mathcal{F}_{\textnormal{Rx},(b)}\left(W_{1,c,(b-1)}, {W}_{2,c,(b-1)}, W_{1,p,(b)}, W_{2,p,(b)}, J_{1,b-1}^*, J_{2,(b-1)}^*\right)   \; \Bigg| \; \bigcup_{b'=1}^{B+1} \left\{ \bar{ \mathcal{E}}_{\textnormal{Tx},1,(b')}, \; \bar{ \mathcal{E}}_{\textnormal{Tx},2,(b')} \right\}\
				\Bigg)   \IEEEeqnarraynumspace\\
				& \leq & \sum_{\substack{(w_{1,c}, w_{2,c}, w_{1,p}, w_{2,p}, j_1, j_2)   \neq \\(W_{1,c,(b-1)}, {W}_{2,c,(b-1)}, W_{1,p,(b)}, W_{2,p,(b)}, J_{1,b-1}^*, J_{2,(b-1)}^*)}  }\Pr \Bigg(  \mathcal{F}_{\textnormal{Rx},(b)}(w_{1,c}, w_{2,c}, w_{1,p}, w_{2,p}, j_1, j_2)    \; \Bigg| \; \bigcup_{b'=1}^{B+1} \left\{ \bar{ \mathcal{E}}_{\textnormal{Tx},1,(b')}, \; \bar{ \mathcal{E}}_{\textnormal{Tx},2,(b')} \right\}\
				\Bigg) 
				\nonumber \\
				& & \hspace{1cm} + \quad \Pr\left( \mathcal{F}_{\textnormal{Rx},(b)}\left(W_{1,c,(b-1)}, {W}_{2,c,(b-1)}, W_{1,p,(b)}, W_{2,p,(b)}, J_{1,b-1}^*, J_{2,(b-1)}^*\right)   \; \Bigg| \; \bigcup_{b'=1}^{B+1} \left\{ \bar{ \mathcal{E}}_{\textnormal{Tx},1,(b')}, \; \bar{ \mathcal{E}}_{\textnormal{Tx},2,(b')} \right\}\right)
			\end{IEEEeqnarray}
		\end{subequations}
		\hrule
	\end{figure*}
	where the inequalities hold by the union bound.
	
	By the event in the conditioning and the way we construct the codebooks,  and  by the weak law of large numbers and the Covering Lemma, both summands tend to 0 as $N\to \infty$ if  Conditions \eqref{eq:Rkp}--\eqref{eq:R1R2R2v} hold. 
	
	The scheme satisfies the distortion constraints \eqref{eq:asymptotics_dis} because of \eqref{Th:distortion} and by the weak law of large numbers.
	\section{Fourier-Motzkin Elimination}\label{app:FME}
	We apply the Fourier-Motzkin Elimination Algorithm to show that Constraints~\eqref{eq:subrates} are equivalent to Constraints \eqref{eq:inner} in Theorem~\ref{Th:achievability}.  For ease of notation, we define
	\begin{subequations}
		\mw{\begin{IEEEeqnarray}{rCl}
				I_0 & :=& I( V_1; X_1X_2Y \mid \underline{U}) 	+ I( V_2; X_1X_2YV_1 \mid \underline{U})	\\
				I_1 &:=&  I(V_1;X_1Z_1\mid \underline{U}) \\
				I_2 &:=& I(V_2;X_2Z_2\mid \underline{U}) \\
				I_3&:=&I(U_1;X_2Z_2\mid U_0U_2) \\
				I_4 &:=& I(U_2;X_1Z_1\mid U_0U_1)\\
				I_5&:=&I(V_1;X_2Z_2\mid \underline{U})\\
				I_6 &:=& I(V_2;X_1Z_1\mid \underline{U})\\
				I_7 &:=& I(X_1 X_2; YV_1V_2\mid  \underline{U})\\
				I_8 &:=& I(X_1 ; YV_1V_2\mid \underline{U} X_2 )\\
				I_9  &:=&I(X_2 ; YV_1V_2\mid \underline{U}X_1)
				\\
				I_{10}&:=& I(X_1; Y \mid U_0 X_2 ) 	\\	
				I_{11}&:=&I(X_2; Y \mid U_0 X_1) 		\\	
				I_{12}&:=&
				I(X_1X_2; Y \mid U_0 U_2 ) \\
				I_{13}&:=&
				I(X_1X_2; Y \mid U_0 U_1) 		\\
				I_{14}&:=&
				I(X_1X_2; Y \mid U_0  ) \\
				I_{15}&:=&I(X_1X_2; Y ).
		\end{IEEEeqnarray}}
	\end{subequations}
	Setting $R_{k,c}=R_k-R_{k,p}$, which is obtained from \eqref{eq:d1}, with above definitions we can rewrite Constraints \eqref{eq:subrates} as: 
	\begin{subequations}\begin{IEEEeqnarray}{rCl}
			R_{1,v} &>& I_1 \label{eq:R1vfme}\\
			R_{2,v}&>&I_2\label{eq:R2v}\\ 
			\mw{R_{2,v}+}R_{1}-R_{1,p}&<&I_2+ I_3\label{eq:R1c}\\
			\mw{R_{1,v}+}R_{2}-R_{2,p}&<&  I_1+I_4
			\label{eq:R2cfme}\\
			R_{1,v}+\mw{R_{2,v}+}R_{1}-R_{1,p}&<&I_2+ I_3+I_5
			\label{eq:R1vR1c}
			\\
			\mw{R_{1,v}+}R_{2,v}+R_{2}-R_{2,p}&<& I_1+I_4+I_6
			\label{eq:R2vR2c}\\
			R_{1,p}+R_{2,p} &<& I_7
			\label{eq:R1pR2p_FME}
			\\
			R_{1,p} &< & I_8
			\\
			R_{2,p}& < & I_9
			\\
			R_{1,v}+R_{1,p}&<& I_{10}+I_0		
			\\	
			R_{2,v}+R_{2,p}&<& I_{11}+I_0
			\\	
			R_{1,v}+R_{1,p}+R_{2,p}&<&
			I_{12}+I_0\\ 
			R_{2,v}+R_{1,p}+R_{2,p}&<& I_{13}+I_0
			\\
			R_{1,v} +R_{1,p}+ R_{2,v}+R_{2,p}&<&
			I_{14}+I_0
			\\
			R_{1,v}+R_1 +R_{2,v}+R_2 & <&  I_{15}+I_0.\label{FME1}
		\end{IEEEeqnarray}
	\end{subequations}
	In a next step we eliminate the variables $R_{1,v}$ and $R_{2,v}$ to obtain:
	\begin{subequations}\begin{IEEEeqnarray}{rCl}
			R_1-R_{1,p} &<&I_3\label{eq:d2}\\
			R_2-R_{2,p}&<&I_4\label{eq:d3}\\ 
			R_1-R_{1,p}&<&I_3+I_5-I_1\label{eq:d4}\\
			R_2-R_{2,p}&<& I_4+I_6-I_2\label{eq:d5}\\
			R_{1,p} &<&\min\{  I_8,  I_{10}+I_0-I_1\}\\
			R_{2,p} &<&\min\{ I_9, I_{11}+I_{0}-I_2\}\\
			R_{1,p}+R_{2,p}&<&\min\{I_7, I_{12}+I_0-I_1,\nonumber \\
			&& \qquad I_{13}+I_0-I_2, I_{14}+I_0-I_1-I_2\} \IEEEeqnarraynumspace\\
			R_{1}+R_{2}&<&I_{15}+I_0-I_1-I_2
		\end{IEEEeqnarray}
	\end{subequations}
	Notice that $I_{1} \geq I_5$ and $I_{2} \geq I_6$ because $V_1-(Z_1, X_1, \underline{U})-(X_2, Z_2)$ form a Markov chain, and thus Constraints \eqref{eq:d2} and \eqref{eq:d3} are inactive in view of Constraints \eqref{eq:d4} and \eqref{eq:d5}. We  thus neglect \eqref{eq:d2} and \eqref{eq:d3} in the following. Eliminating next variable $R_{1,p}$, where we take into account the nonnegativity of $R_{1,p}$ and $R_{1}-R_{1,p}$, we obtain: 
	\begin{subequations}\begin{IEEEeqnarray}{rCl}
			R_1 &<&I_3 +I_5-I_1+\min\{  I_8,  I_{10}+I_0-I_1\}\IEEEeqnarraynumspace\\
			R_1 +R_{2,p} &<&I_3+I_5-I_1 + \min\{I_7, I_{12}+I_0-I_1,\nonumber \\
			&& \qquad I_{13}+I_0-I_2, I_{14}+I_0-I_1-I_2\} \IEEEeqnarraynumspace\\
			R_2-R_{2,p}&<& I_4+I_6-I_2\\
			R_{2,p} &<&\min\{ I_9, I_{11}+I_{0}-I_2\} \label{eq:f1}\\
			R_{2,p}&<&\min\{I_7, I_{12}+I_0-I_1,\nonumber \\
			&& \qquad I_{13}+I_0-I_2, I_{14}+I_0-I_1-I_2\}\label{eq:f2} \IEEEeqnarraynumspace\\
			R_{1}+R_{2}&<&I_{15}+I_0-I_1-I_2
		\end{IEEEeqnarray}
		and 
		\begin{IEEEeqnarray}{rCl}
			I_3+I_5&>& I_1\\
			I_{10}+I_0 &>& I_1.
		\end{IEEEeqnarray}
	\end{subequations}
	Notice that $I_7 > I_9$ and $I_{13}>I_{11}$ and therefore the two Constraints \eqref{eq:f1} and \eqref{eq:f2} combine to 
	\begin{IEEEeqnarray}{rCl}
		R_{2,p} &<&\min\{ I_9, I_{11}+I_{0}-I_2, \nonumber \\
		&& \qquad I_{12}+I_0-I_1,I_{14}+I_0-I_1-I_2\}.
	\end{IEEEeqnarray}
	Eliminating finally 
	$R_{2,p}$ (while taking into account the nonnegativity of $R_{2,p}$ and $R_{2}-R_{2,p}$) results in: 
	\begin{subequations}
		\begin{IEEEeqnarray}{rCl}
			R_1&<&I_3+I_5-I_1+\min\{ I_8,\;  I_{10}+I_0-I_1\} \label{eq:fff}\IEEEeqnarraynumspace \\
			R_1 &<&I_3+I_5-I_1 + \min\{I_7, I_{12}+I_0-I_1,\nonumber \\
			&& \qquad I_{13}+I_0-I_2, I_{14}+I_0-I_1-I_2\}\label{eq:fffDk}
			\\
			%
			R_2&<& I_4+I_6-I_2+\min\{I_9, \;  I_{11}+I_{0}-I_2 \nonumber \\
			&& \qquad I_{12}+I_0-I_1,I_{14}+I_0-I_1-I_2\}\\
			R_1+R_2&<& I_4+I_6- I_2+I_3+I_5- I_1\nonumber\\
			&&\hspace{0cm}+	\min\{I_7, \; I_{12}+I_0-I_1,
			\nonumber\\&&\hspace{1cm}I_{13}+I_0-I_2,I_{14}+I_0-I_1-I_2
			\} 	\\
			R_{1}+R_{2}&<&I_{15}+I_0-I_1-I_2
		\end{IEEEeqnarray}
		and 
		\begin{IEEEeqnarray}{rCl}
			I_3+I_5 &>& I_1\\
			I_4 +I_6 & >& I_2\\
			I_{14}+I_0& >& I_1 +I_2\\
			I_{10}+I_0 &>& I_1\label{eq:d0}\\
			I_{11} +I_0 & > & I_2\\
			I_{12} +I_0 & > & I_1.\label{eq:d}
		\end{IEEEeqnarray}
	\end{subequations}
	Notice that $I_{12}>I_{10}$ and thus \eqref{eq:d} is obsolete \mw{in view of \eqref{eq:d0}. Moreover, since} also $I_7>I_8$, Constraints \eqref{eq:fff} and \eqref{eq:fffDk} combine to 
	\begin{IEEEeqnarray}{rCl}
		R_1&<&I_3+I_5-I_1+\min\{ I_8,\;  I_{10}+I_0-I_1,\nonumber \\
		&& \qquad I_{13}+I_0-I_2, I_{14}+I_0-I_1-I_2\} .\IEEEeqnarraynumspace
	\end{IEEEeqnarray}
	The final expression is thus given by constraints: 
	\begin{subequations}
		\begin{IEEEeqnarray}{rCl}
			R_1&<&I_3+I_5-I_1+\min\{ I_8,\;  I_{10}+I_0-I_1\nonumber \\
			&& \qquad I_{13}+I_0-I_2, I_{14}+I_0-I_1-I_2\} \IEEEeqnarraynumspace\\
			R_2&<& I_4+I_6-I_2+\min\{I_9, \;  I_{11}+I_{0}-I_2 \nonumber \\
			&& \qquad I_{12}+I_0-I_1,I_{14}+I_0-I_1-I_2\}\\
			R_1+R_2&<& I_4+I_6- I_2+I_3+I_5- I_1\nonumber\\
			&&\hspace{0cm}+	\min\{I_7, \; I_{12}+I_0-I_1,
			\nonumber\\&&\hspace{1cm}I_{13}+I_0-I_2,I_{14}+I_0-I_1-I_2
			\} 	\\
			R_{1}+R_{2}&<&I_{15}+I_0-I_1-I_2
		\end{IEEEeqnarray}
		and 
		\begin{IEEEeqnarray}{rCl}
			I_3+I_5 &>& I_1\\
			I_4 +I_6 & >& I_2\\
			I_{14}+I_0& >& I_1 +I_2\\
			I_{10}+I_0 &>& I_1\\
			I_{11} +I_0 & > & I_2.
		\end{IEEEeqnarray}
	\end{subequations}
	\bibliographystyle{IEEEtran}
	\bibliography{JRC,main_v2}	
\end{document}